\documentclass[apj]{emulateapj}
\usepackage[breaklinks,colorlinks,citecolor=blue]{hyperref}
\usepackage{graphicx}

\received{September 22, 2016}
\revised{March 27, 2017}
\accepted{April 5, 2017}

\shorttitle{ALMA Observations of d216-0939}
\shortauthors{Factor et al.}

\begin{document}

\title{ALMA Observations of Asymmetric Molecular Gas Emission from a Protoplanetary Disk in the Orion Nebula}


\author{Samuel M. Factor\altaffilmark{1,2}}
\author{A.M. Hughes\altaffilmark{1}}
\author{Kevin M. Flaherty\altaffilmark{1}}
\author{Rita K. Mann\altaffilmark{3}}
\author{James Di Francesco\altaffilmark{3,4}}
\author{Jonathan P. Williams\altaffilmark{5}}
\author{Luca Ricci\altaffilmark{6}}
\author{Brenda C. Matthews\altaffilmark{3,4}}
\author{John Bally\altaffilmark{7}}
\author{Doug Johnstone\altaffilmark{3,4}}

\altaffiltext{1}{Department of Astronomy, Van Vleck Observatory, Wesleyan University, 96 Foss Hill Drive, Middletown, CT 06459, USA}
\altaffiltext{2}{Department of Astronomy, The University of Texas at Austin, TX 78712, USA}
\altaffiltext{3}{National Research Council Canada Herzberg Astronomy and Astrophysics, 5071 West Saanich Road, Victoria, BC, V9E 2E7, Canada} 
\altaffiltext{4}{Department of Physics and Astronomy, University of Victoria, Victoria, BC, V8P 1A1, Canada } 
 \altaffiltext{5}{Institute for Astronomy, University of Hawaii, 2680 Woodlawn Drive, Honolulu, HI 96822, USA}
 \altaffiltext{6}{Harvard-Smithsonian Center for Astrophysics, 60 Garden Street, Cambridge, MA 02138, USA}
 \altaffiltext{7}{CASA, University of Colorado, CB 389, Boulder, CO 80309, USA} 

\email{sfactor@astro.as.utexas.edu}

\begin{abstract}
We present Atacama Large Millimeter/submillimeter Array (ALMA) observations of molecular line emission from d216-0939, one of the largest and most massive protoplanetary disks in the Orion Nebula Cluster (ONC). We model the spectrally resolved HCO$^+$ (4--3), CO (3--2), and HCN (4--3) lines observed at 0\farcs5 resolution to fit the temperature and density structure of the disk. We also weakly detect and spectrally resolve the CS (7--6) line but do not model it. The abundances we derive for CO and HCO$^+$ are generally consistent with expected values from chemical modeling of protoplanetary disks, while the HCN abundance is higher than expected. We dynamically measure the mass of the central star to be $2.17\pm0.07\,M_\sun$ which is inconsistent with the previously determined spectral type of K5. We also report the detection of a spatially unresolved high-velocity blue-shifted excess emission feature with a measurable positional offset from the central star, consistent with a Keplerian orbit at $60\pm20$\,au. Using the integrated flux of the feature in HCO$^+$ (4--3), we estimate the total H$_2$ gas mass of this feature to be at least $1.8-8\,M_\mathrm{Jupiter}$, depending on the assumed temperature. The feature is due to a local temperature and/or density enhancement consistent with either a hydrodynamic vortex or the expected signature of the envelope of a forming protoplanet within the disk. 
\end{abstract}


\section{Introduction}
\label{intro}

Planetary systems are born from the gas-rich ``protoplanetary'' disks commonly observed around young stars \citep{diskFir,diskFsmm,diskF}. The properties of these disks provide insight into the conditions under which planets form and the physics underlying the process of planet formation. Protoplanetary disks tend to have a flared structure with radii up to a few hundred au \citep{size}, temperatures in the outer disk between 10--100 K, and disk masses from $\lesssim0.001\,M_\odot$ to $\sim0.1\,M_\odot$ \citep[$\lesssim1-100\,M_\mathrm{Jupiter}$,][]{SMTauAur,Ansdell2016}. Optically thick dust disks tend to live for a few Myr \citep[$\lesssim10$,][]{Dages} after which the primordial dust and gas disperses and a tenuous, dusty debris disk may emerge \citep{review}. 

Submillimeter observations make it possible to study the interior regions of protoplanetary disks, close to the midplane, which are hidden under the optically thick disk surface at shorter wavelengths. Optically thin continuum emission provides a measurement of the underlying mass distribution of solids, while molecular lines provide insight into the chemistry, kinematics, temperature, and density of the gas in the disk. Optically thin molecular lines exhibit a degeneracy between temperature and density, while optically thick lines probe the temperature at the $\tau=1$ surface layer of a particular species. With a combination of optically thick and thin molecular line modeling it is possible to constrain both the temperature and density structure of the disk \citep[see, e.g.,][]{pie07,pan08,qi11,weakTurb}. To estimate the total gas mass (dominated by H$_2$ which is effectively invisible at the low temperatures in the outer disk), it is necessary to assume a fractional abundance of the gas species observed. The typically assumed fractional abundance of CO is $10^{-4}$ \citep[e.g.][]{abund,abund2}, although there is evidence that it may deviate from this value in protoplanetary disks \citep[e.g.,][]{gasToDust,TWhHD}. For other species that exhibit more complicated chemistry, ISM-based values or those from chemical simulations of disks may be used \citep[e.g.][]{HCOpCle,ionHCN}.

Spatially and spectrally resolved observations and subsequent modeling of nearby circumstellar disks have permitted characterization of many aspects of their structure. Observations of dust continuum emission from disks in nearby low-mass star-forming regions \citep[SFRs;][]{SMTauAur,SMOph,OphStruct,OphStruct2,HighResCARMA} have yielded a wide distribution of disk masses with a median of approximately 0.005 $M_{\odot}$. A large fraction of the surveyed disks are more massive than the minimum mass solar nebula \citep[MMSN;][]{MMSN} of roughly 0.01 $M_{\odot}$, implying that the potential to form solar systems similar to our own is common. Most stars \citep[probably including the Sun; see, e.g.,][]{Fe,Al} however, form in massive, dense clusters known as high-mass SFRs \citep[e.g.,][]{clusters}. These environments are extraordinarily different from the low-mass SFRs where most of the disk observations to date have taken place. These regions host more high-mass O and B stars, which give off ionizing radiation, and a higher density of stars, increasing the chance of gravitational interaction. Both of these features of high-mass SFRs cause mass loss in protoplanetary disks, which could hinder planet formation potential by altering the temperature and density structure of the protoplanetary disk. 

The Orion Nebula Cluster (ONC) is the closest high-mass SFR. {\it Hubble Space Telescope} observations of several protoplanetary disks in the central regions of the ONC demonstrate that they are surrounded by tear-drop-shaped shells pointing away from $\theta^1$ Ori C \citep[Spectral type O6,][]{HSTOri,Photoevaporation,HSTOri2,HST,HSTProplyds}. These images beautifully illustrate the harsh environment created by the nearby O star and indicate that it may have an effect on the evolution of protoplanetary disks. Separate observations by \citet{massLoss1}, using the Very Large Array (VLA), and \citet{massLoss2}, using Keck, have measured mass loss rates of $\dot{M} \approx 10^{-7}$ $M_\odot\,\mathrm{yr}^{-1}$. Such a substantial mass loss rate would disperse a typical disk before giant planets could form \citep{planetFormTime}, and is in apparent contradiction with the inferred ages of the disk-hosting ONC stars \citep[$\sim$2 Myr,][]{age1,age2}. Thus, it is important to gain a better understanding of the effects of environment on protoplanetary disks and compare disks in these regions, which are more typical of the general disk population, to disks previously studied in low-mass SFRs. 

All previous submillimeter studies of disk masses in the ONC have used observations of millimeter continuum emission from dust \citep[or the silhouette of an optically thick line against the cloud background; see][]{bal15}. Surveys using the Submillimeter Array (SMA) by \citet{MassiveOrion,SMAOri2,SMAOri} have shown that the upper end of the disk mass distribution is truncated in the ONC and there is a positive correlation of disk mass with distance from $\theta^1$ Ori C. Observations by \citet{ALMAOri} using ALMA have confirmed these results. They attribute this trend to external photoevaporation since stellar interactions remove relatively small amounts of mass. They also note that despite the observed disk destruction, the region shows similar potential for planet-formation to that of low-mass SFRs, with 30\% of surveyed disks having masses greater than or equal to the MMSN. For comparison, $\sim37\%$ and $29\%$ of disks in Taurus and $\rho$ Ophiuchus respectively have derived disk masses greater than or equal to the MMSN \citep{SMTauAur,SMOph}. However, the measurements to date primarily or exclusively rely on dust continuum emission as a probe of total disk mass, and do not investigate the effects of the high-mass star-forming environment on the structure and chemistry of the gas disk. 

Here we present molecular line observations of the disk around d216-0939, a pre-main-sequence star with a spectral type of K5 \citep{SpT} located in the outskirts of the Orion Nebula \citep[projected distance 1.6\,pc;][see their Figure 1]{ALMAOri}. The disk was discovered by \citet{HST} using HST and is one of the largest and most massive disks in the ONC \citep{MassiveOrion,SMAOri,ALMAOri}. \citet{HST} determined that the disk is almost edge-on with an inclination of $\sim75^\circ-80^\circ$ to the east\footnote{This value is derived from the fact that the line of sight to the star is very close to the flared surface of the disk. They report the inclination with respect to the disk axis rather than with respect to the plane of the disk and thus report a value of $\sim10^\circ-15^\circ$.} (based on the morphology of the reflection nebula, the far side of the disk is east of the star while the near side is west) and a polar axis position angle of $83^\circ$ (E of N). They also noted that the northern portion of the disk, as seen in scattered light, is $\sim50\%$ larger than the southern portion. The disk is also associated with HH 667 E, a partial bow shock slightly bent to the south, and HH 667 W, several diffuse filaments along the rotation axis of the disk. 

The dust disk has been well characterized at millimeter wavelengths. Continuum observations using the SMA by \citet{MassiveOrion} have marginally resolved the dust disk and determined its mass to be $0.0450\pm0.0006$ $M_\odot$ with a radius of 291\,au. They detected no signs of external photoevaporation, as would be expected from its location relatively far away from any O or B stars. Using spatially unresolved observations from the Combined Array for Research in Millimeter Astronomy (CARMA) and the Australia Telescope Compact Array (ATCA), \citet{SED} were able to fit the long wavelength spectral index to determine a dust opacity index of $\beta=1.0\pm0.3$ (where the dust opacity $\kappa_\nu\propto\nu^\beta$), indicative of grain growth. They also derive a disk mass of 0.02 $M_\odot$. Analysis of continuum data from ALMA (from which gas line data are presented here) by \citet{ALMAOri} yielded a disk mass of $0.0437\pm0.0007$ $M_\odot$ with a radius of 525\,au. \citet{MassiveOrion} also strongly detected the CO (3--2) transition but were unable to separate the disk from the ambient cloud emission (i.e. cloud contamination). 

We investigate the disk structure of d216-0939 using sensitive ALMA observations of several molecular lines: the higher sensitivity reveals the temperature structure of the inner disk in the wings of the CO line, which we combine with lower-abundance tracers (HCO$^+$ and HCN) that exhibit less cloud contamination than CO and allow us to study the outer gas disk for the first time. We measure the temperature and density structure of the disk, the mass of the host pre-main sequence star, and provide constraints on the molecular abundances. Comparing these values with those previously measured in low-mass SFRs provides insight into the effect of environment on planet-forming potential. The work is organized as follows: The observations and data reduction are described in Section \ref{obs}. The basic observational parameters of the system are presented in Section \ref{res}. Modeling of the temperature and density structure of the disk from the gas lines is discussed and best-fit models are presented in Section \ref{mod}. The implications of the modeling results are discussed in Section \ref{disc}, including the surprising discovery of a high-velocity feature located close to the star that resembles either a localized planet-forming vortex or the molecular envelope of a protoplanet forming within the disk.

\section{Observations}
\label{obs}

Data presented here are part of an ALMA survey of the Orion proplyds (project 2011.0.00028.S). Data collection methods and analysis of the continuum results are presented in \citet{ALMAOri}. The observations were made on October 24, 2012, using the Band 7 (350 GHz, $\sim850\,\mu$m) receivers. Four 1.875 GHz-wide spectral windows were arranged to cover the HCO$^+$ (4--3), HCN (4--3), CO (3--2), and CS (7--6) emission lines at 356.734\,GHz, 354.505\,GHz, 345.796\,GHz, and 342.883\,GHz respectively. Each window was divided into 3840 channels of 488.28\,kHz width, corresponding to a velocity resolution of 0.42\,km\,s$^{-1}$. For this Cycle 0 Early Science project, 22 12-m antennas were online in a hybrid configuration with baselines ranging from 21.2\,m to 384.2\,m. This configuration results in a largest recoverable scale of $8''$ and an angular resolution of 0\farcs5 (3,500 and 190\,au respectively, at 414\,pc, the distance of Orion we chose to adopt in this work).

The distance to the Orion Nebula has been measured with the geometric parallax method using the Very Long Baseline Array \citep[$389^{+24}_{-21}$\,pc, $392\pm34$\,pc, $414\pm7$\,pc;][respectively]{OD1,OD4,OD2} and by orbit monitoring of $\theta^1$ Ori C \citep[$410\pm20$;][]{OD3}. We have used a distance of 414 pc in this work, corresponding to the most recent trigonometric parallax measurement, which provides the most precise measurement of the distance.

During the analysis of CO (3--2) and HCO$^+$ (4--3), baselines shorter than $70\,k\lambda$ were excluded to minimize large-scale cloud contamination, resulting in a largest recoverable scale of 2\farcs9 (1,200\,au; see Section \ref{res} below for details). The data presented here, from Field 5 of \citet{ALMAOri}, represent 22 minutes of on-source time, achieving an rms noise of 0.41\,mJy\,beam$^{-1}$ in the aggregate continuum data and 6\,mJy\,beam$^{-1}$ in each channel. Observations were spaced out over 7.5 hours to ensure adequate $uv$ coverage and resulted in a synthesized beam of $0\farcs57\times0\farcs52$ with a position angle of 88.8$^{\circ}$ using natural weighting. When excluding baselines shorter than $70\,k\lambda$, the synthesized beam was $0\farcs51\times0\farcs47$ with a position angle of 89.5$^\circ$. Precipitable water vapor in the atmosphere was stable at 0.7\,mm.

Data was calibrated by ALMA staff using the standard procedures in the Common Astronomy Software Applications \citep[CASA,][]{CASA} package. The antenna-based complex gains and bandpass response of the system were calibrated using observations of the quasars J0607-085 and J0522-364 respectively. The absolute flux calibration was determined from observations of Callisto. The model of Callisto was that provided by \citet{Callisto}. Absolute flux calibration is estimated to be accurate to within $\sim10\%$ \citep{ALMAOri}.

The velocity reference frame was converted from the ALMA standard topocentric to LSRK (kinematical local standard-of-rest) using the CASA task \texttt{cvel}, and continuum subtraction in the uv plane was performed using the CASA task \texttt{uvcontsub}. Visibilities were then inverted with natural weighting, deconvolved, and restored using the standard procedures in the Multichannel Image Reconstruction Image Analysis and Display \citep[MIRIAD,][]{miriad} package.

\section{Results}
\label{res}

Spatially and spectrally resolved line emission was detected for CO (3--2), HCO$^+$ (4--3), HCN (4--3), and CS (7--6) across approximately 40 channels with a channel width of 0.42\,km\,s$^{-1}$. Detailed studies of gas structure in protoplanetary disks are still relatively rare, and have previously only focused on nearby low-mass SFRs. Thus, we would like to gain a general understanding of the observations before beginning detailed modeling. This includes removing cloud contamination, investigating the general morphology of the disk using moment maps, estimating the gas mass using the integrated line flux, and examining the velocity profile to estimate the mass of the central star. We also detect high-velocity blue-shifted excess emission near the central star and a possible outflow feature.

Cloud contamination occurs when emission from background or foreground gas clouds is detected in the same direction as the circumstellar disk. Since the Orion Nebula is a denser region of space than the low-mass SFRs of Taurus and Ophiuchus, cloud contamination is much more common and problematic. Previous observations of d216-0939 using the SMA by \cite{MassiveOrion} strongly detected the CO (3--2) transition but the disk could not be distinguished from the contamination. Due to the higher sensitivity of the ALMA observations, we can detect some CO (3--2) emission from the disk in the high-velocity wings of the line, for which the velocity offset minimizes cloud contamination. For the other molecular lines, which have higher critical densities, cloud contamination is less severe but still clearly present in HCO$^+$, which is the brightest tracer. Since cloud contamination tends to be large scale in nature, excluding short baselines reduces its contribution to the observations. While this process slightly reduces the total recovered flux from the disk and our ability to characterize its large-scale structure, it is necessary to avoid including emission from the cloud in the fits to the disk emission. Since HCO$^+$ is the most contaminated line after CO, we used the HCO$^+$ data to select the optimal range of baseline lengths to include in the fit to the data. We found that excluding baselines shorter than $70\,k\lambda$ (corresponding to an angular scale of $\sim3''$) best minimized the intensity of the cloud contamination feature located to the southwest of the disk while only marginally reducing the flux recovered from the disk. 

\begin{figure*}[htb!]
\begin{center}
    \includegraphics[trim=0cm 4.1cm 0cm 3cm, clip=true, width=0.49\textwidth]{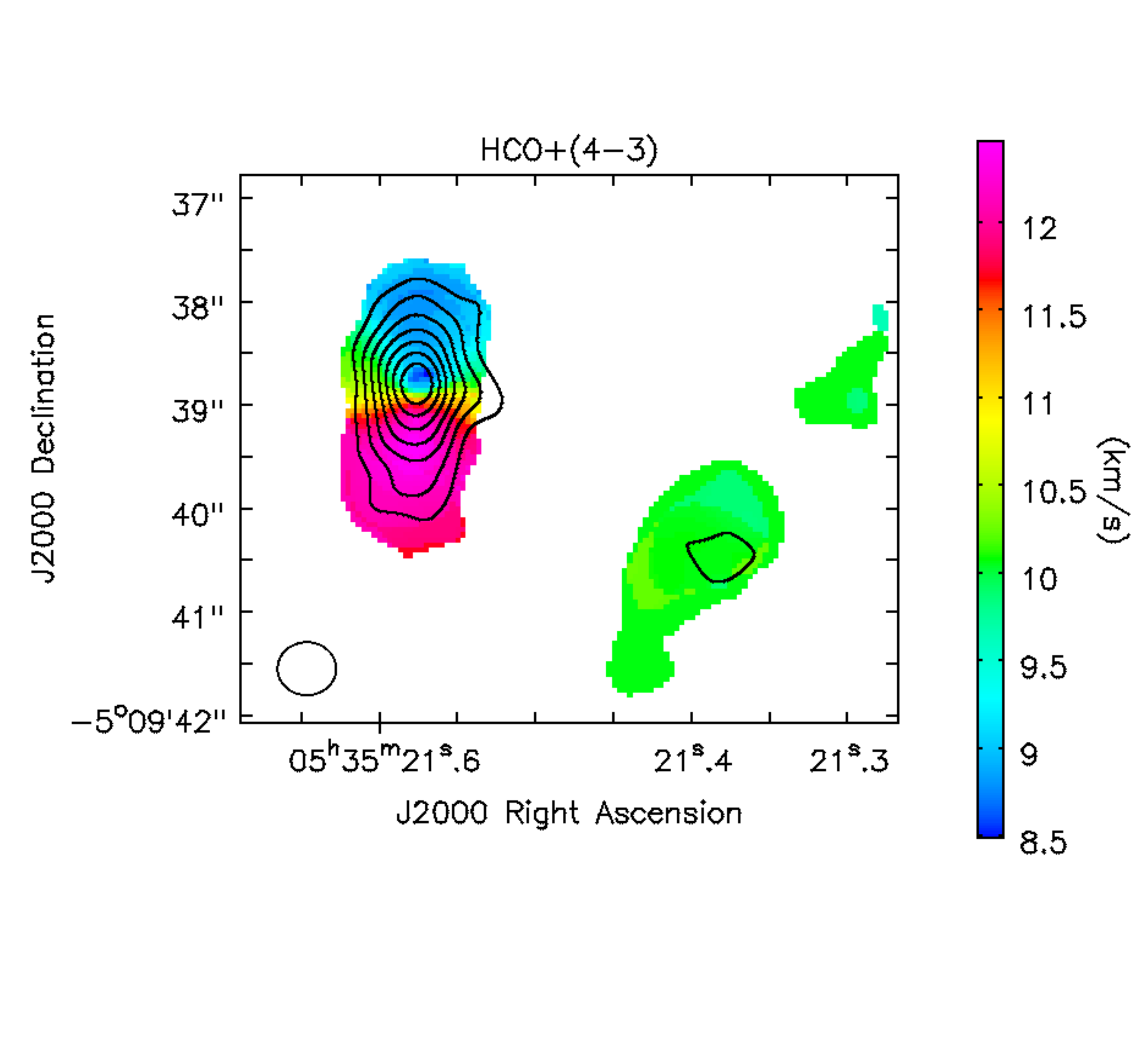}
    \includegraphics[trim=0cm 4.1cm 0cm 3cm, clip=true,width=0.49\textwidth]{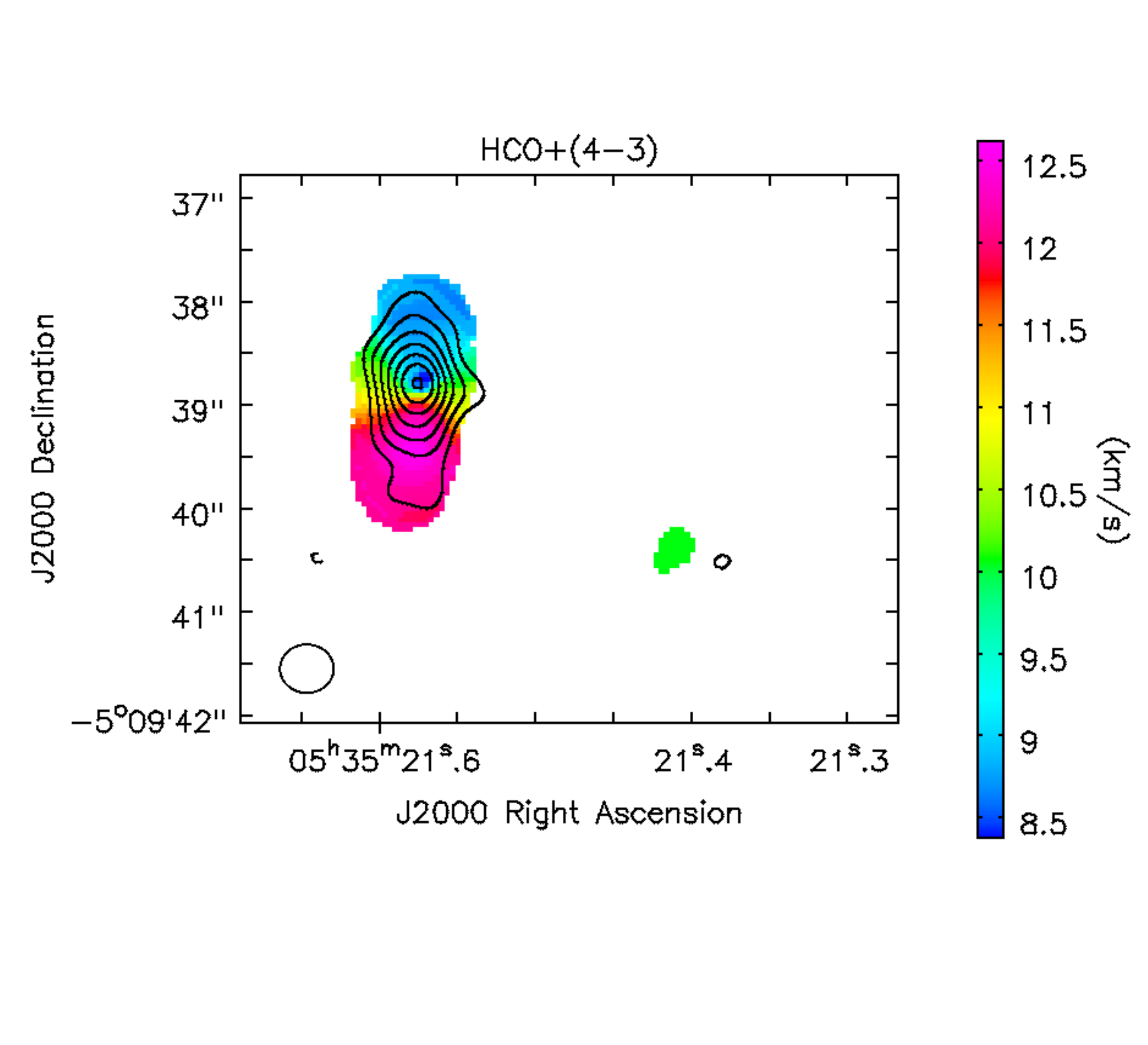}
    \caption{Integrated intensity maps of the HCO$^+$ (4--3) transitions including all baselines (left) and excluding baselines shorter than $70\,k\lambda$ (right). Excluding short baselines clearly reduces cloud contamination (i.e. the feature in the south-west). Contours are integrated line intensity at 3, 5, 7\ldots15\,$\sigma$ where $1\,\sigma$ is 32\,mJy\,beam$^{-1}$ km s$^{-1}$ and 28\,mJy\,beam$^{-1}$ km s$^{-1}$ respectively. Colors show the intensity-weighted velocity (LSRK). The FWHM size of the synthesized beam is shown in the bottom left corner. The beam diameter is 0\farcs5, corresponding to 207\,au at this distance.}
\label{fig:HCOpmoms}
\end{center}
\end{figure*}

\begin{figure*}[htb!]
\begin{center}
    \includegraphics[trim=0cm 4.2cm 0cm 3cm, clip=true,width=0.49\textwidth]{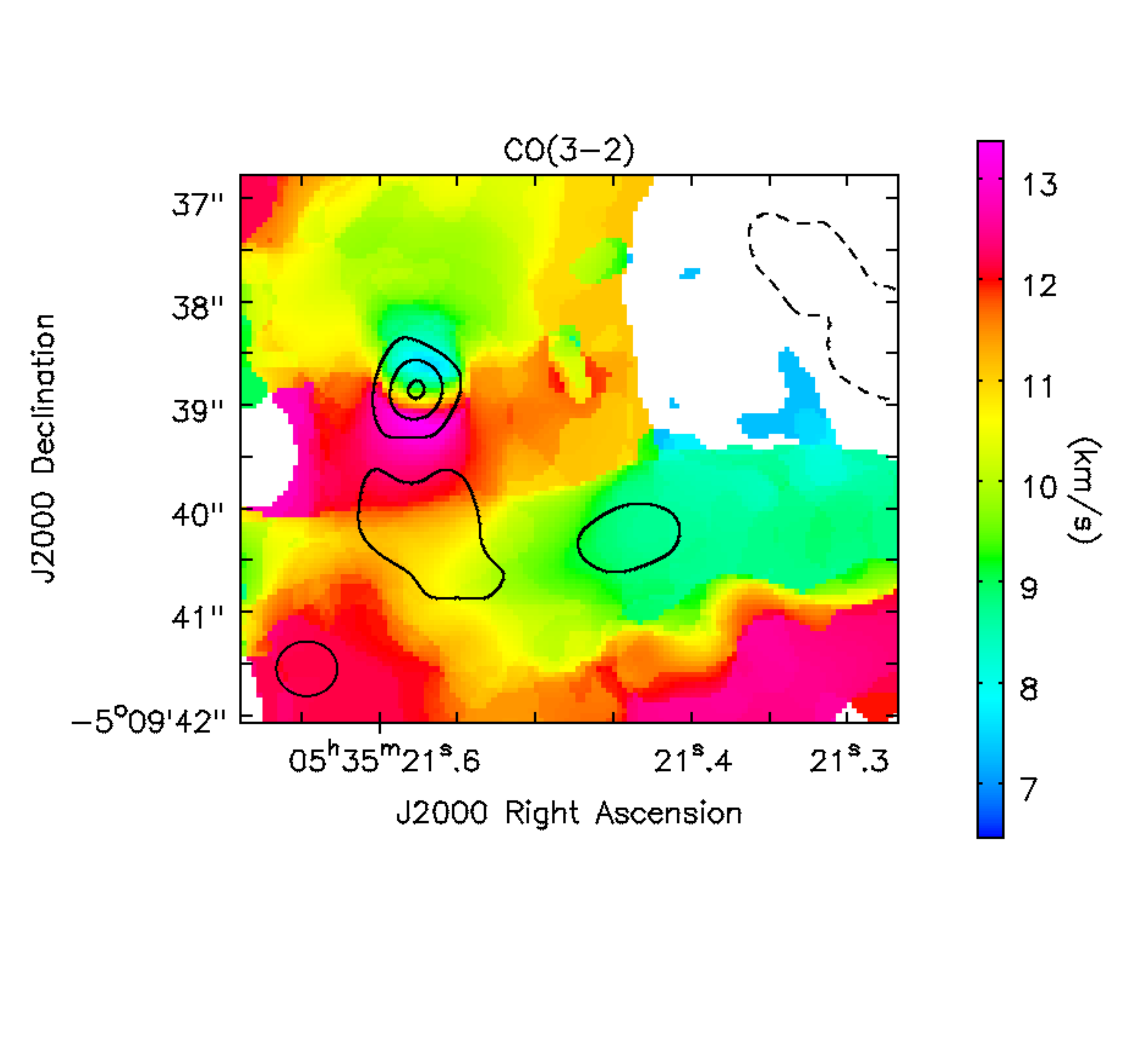}
    \includegraphics[trim=0cm 4.2cm 0cm 3cm, clip=true,width=0.49\textwidth]{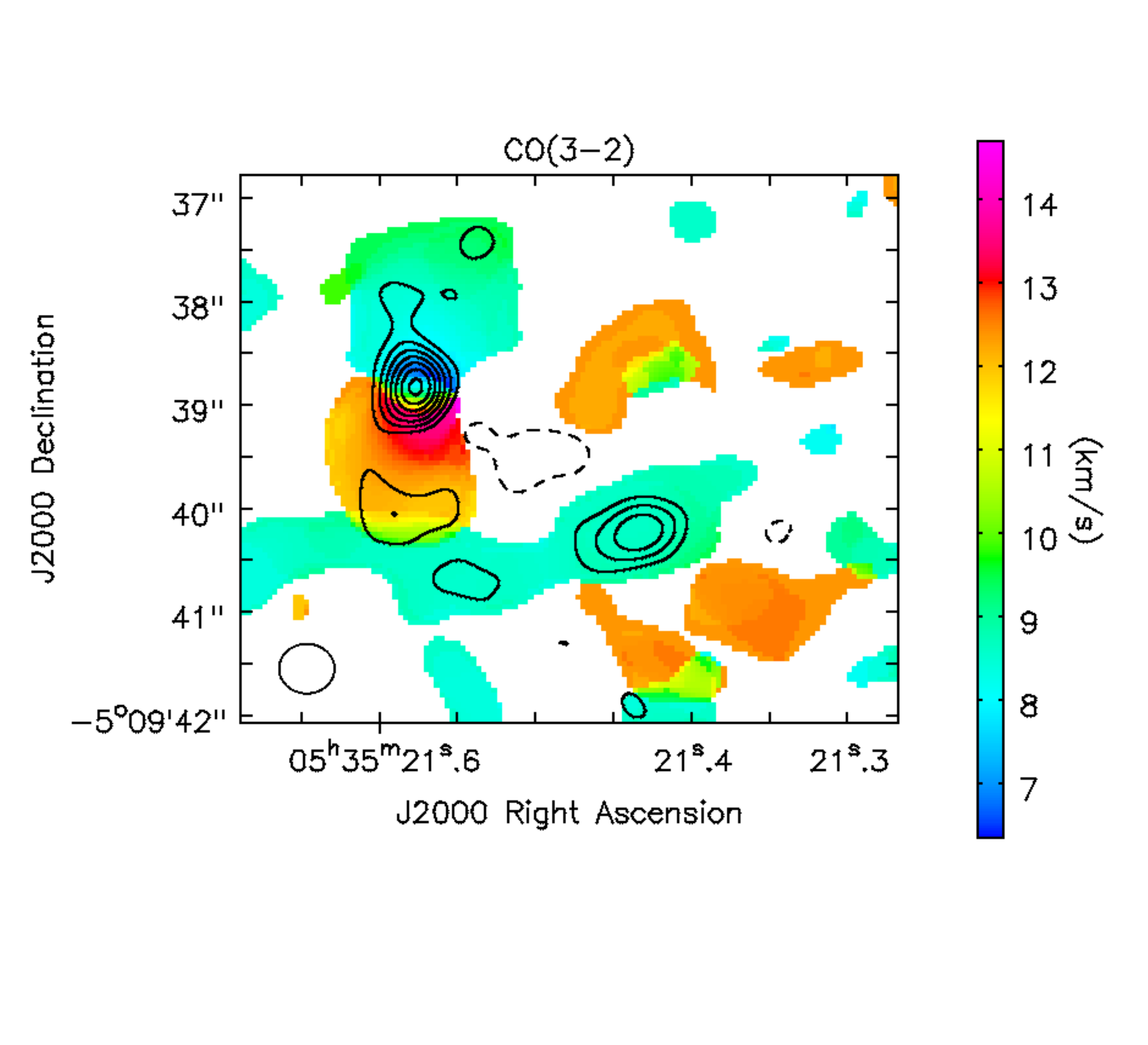}
    \caption{Integrated intensity maps of the CO (3--2) transition including all baselines (left) and excluding all baselines shorter than $70\,k\lambda$ (right). Excluding short baselines clearly reduces, though does not eliminate cloud contamination. Black contours are integrated line emission at $\pm3$, 5, 7\ldots15\,$\sigma$ where $1\,\sigma$ is 132\,mJy\,beam$^{-1}$ km s$^{-1}$ and 47\,mJy\,beam$^{-1}$ km s$^{-1}$, respectively. Negative contours are dashed. Colors show the intensity-weighted velocity (LSRK). The FWHM size of the synthesized beam is shown in the bottom left corner. The beam diameter is 0\farcs5, corresponding to 207\,au at this distance.}
\label{fig:COmoms}
\end{center}
\end{figure*}

\begin{figure*}
\begin{center}
    \includegraphics[trim=0cm 3cm 0cm 2.2cm, clip=true,width=0.49\textwidth]{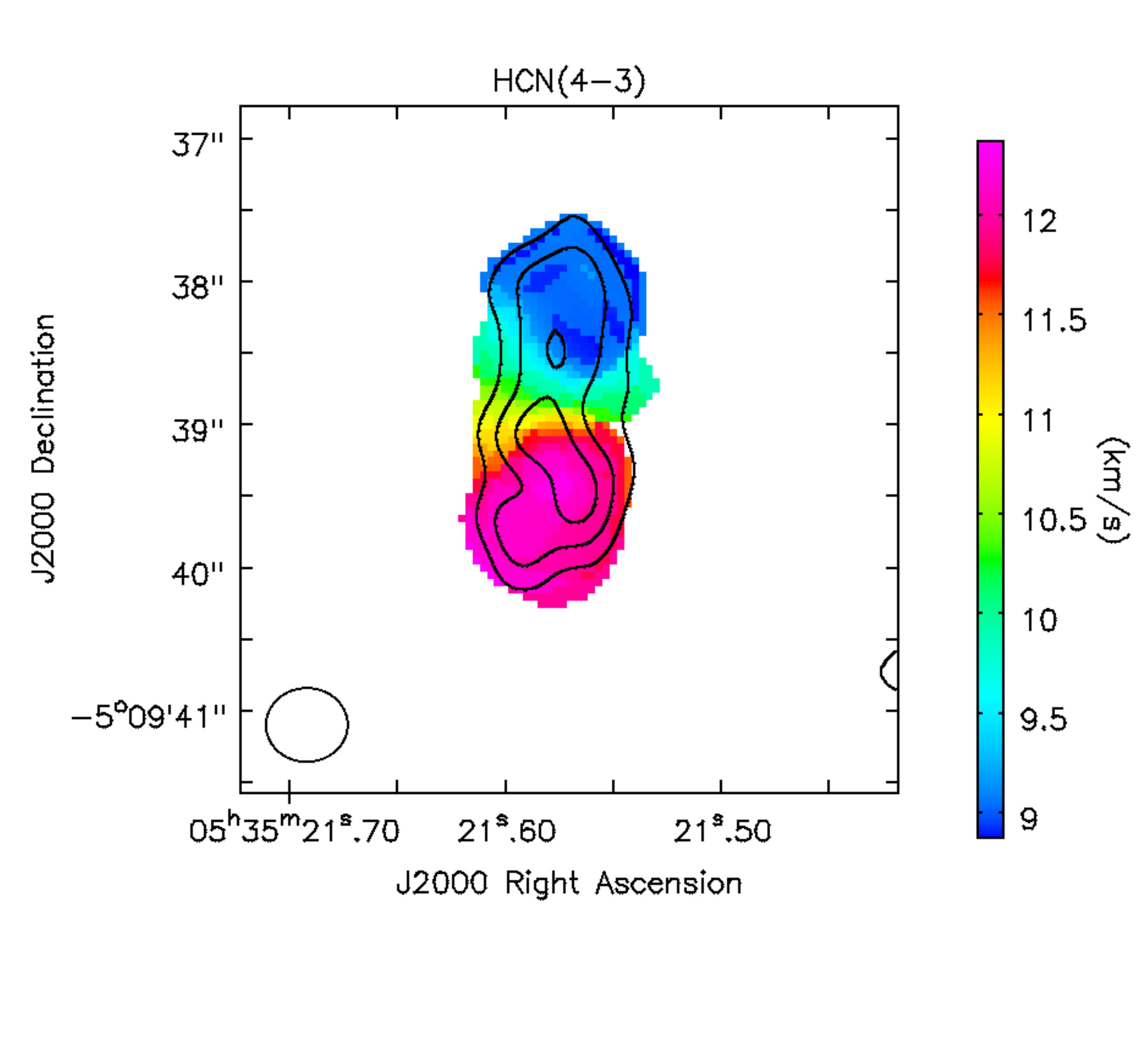}
    \includegraphics[trim=0cm 3cm 0cm 2.2cm, clip=true,width=0.49\textwidth]{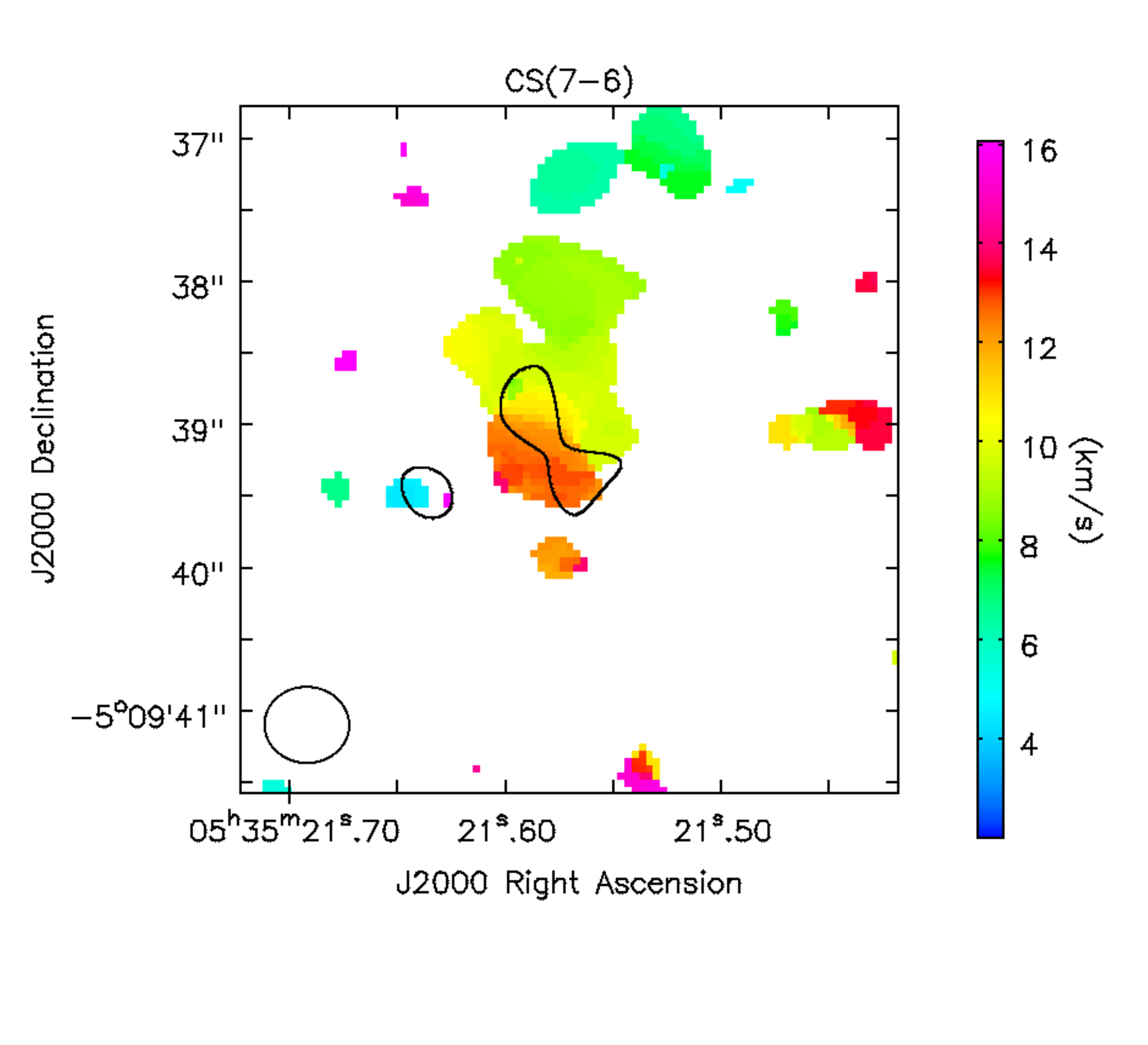}
    \caption{Integrated intensity maps of the HCN (4--3) (left) and CS (7--6) (right) transitions. Black contours are integrated line emission at 3, 5, and 7\,$\sigma$ where $1\,\sigma$ is 23\,mJy\,beam$^{-1}$ km s$^{-1}$ and 21\,mJy\,beam$^{-1}$ km s$^{-1}$, respectively. Colors show the intensity-weighted velocity (LSRK). The FWHM size of the synthesized beam is shown in the bottom left corner. The beam diameter is 0\farcs5, corresponding to 207\,au at this distance.}
\label{fig:moms}
\end{center}
\end{figure*}

The effect of the choice of the exclusion of baselines shorter than 70\,k$\lambda$ is illustrated in Figures~\ref{fig:HCOpmoms} and \ref{fig:COmoms}, which show the integrated intensity (moment 0) and the intensity-weighted velocity (moment 1) maps for HCO$^+$ (4--3) and CO (3--2), respectively. While the zeroth-moment map of CO (3--2) emission still appears highly contaminated, excluding short baselines reduces contamination in some of the individual channels and allows us to use a wider range of channels to characterize the disk structure. The HCO$^+$ (4--3) line is almost completely free of contamination after excluding short baselines. Figure \ref{fig:moms} shows moment maps for the HCN (4--3) and CS (7--6) emission. Both HCN and CS are free of contamination without excluding short baselines. CS emission is detected at the $3\,\sigma$ level and only marginally resolved. In all lines, rotation of the disk is clearly visible as a transition from red-shifted emission in the south to blue-shifted emission in the north. The maximum extent of the $3\,\sigma$ contour along the disk major axis corresponds to an outer diameter of the HCO$^+$ and HCN disks of 950\,au and 1090\,au, respectively at a distance of 414\,pc. The CO emission is too contaminated to provide a good measurement of the extent of the outer disk, and the CS emission is not sufficiently strongly detected to provide a reliable measurement.

Tables \ref{tab:lineF} and \ref{tab:lineGaus} present the velocity-integrated line fluxes and the best-fit parameters for a simple elliptical Gaussian fit to the visibilities, respectively. The integrated line flux was measured using the MIRIAD task \texttt{cgcurs} to integrate the intensity in the zeroth-moment map throughout the region enclosed by the $3\,\sigma$ contour level (noise levels are given in the captions to Figures \ref{fig:HCOpmoms}--\ref{fig:moms}). Stated uncertainties on the integrated line flux do not include the absolute flux calibration uncertainty of the ALMA observations of $\sim10\%$ \citep{ALMAOri} caused by uncertainties in the models of solar system objects used as flux calibrators. Elliptical Gaussian fits to the visibilities were performed using the MIRIAD task \texttt{uvfit}. Channel maps for the CO (3--2) and HCO$^+$ (4--3) emission excluding baselines shorter than 70 k$\lambda$ are presented in Section \ref{sec:fitproc}. 

\begin{deluxetable}{llc}
    \tablewidth{0.865\columnwidth}
    \tablecaption{Integrated Flux Measurements\label{tab:lineF}}
    \tablehead{
    \colhead{Line} & \colhead{Baselines} & \colhead{Integrated Line Flux}\\ & & \colhead{(Jy\,km\,s$^{-1}$)}
    }
\startdata
CS (7--6)      & All           & $0.085\pm0.007$           \\
HCN (4--3)     & All           & $0.84\pm0.03$             \\
HCO$^+$ (4--3) & All           & $1.8\pm0.1$               \\
HCO$^+$ (4--3) & $>70\,k\lambda$ & $1.06\pm0.09$     
\enddata
\tablecomments{Integrated line flux was not calculated for CO (3--2) as the data were too contaminated to give a meaningful result. Uncertainties on the integrated line flux do not include the absolute flux calibration uncertainty of the ALMA observations of $\sim10\%$ \citep{ALMAOri}.}
\end{deluxetable}

\begin{deluxetable*}{llcccc}
    \tablewidth{0.85\textwidth}
    \tablecaption{Gaussian Fits to Visibilities\label{tab:lineGaus}}
    \tablehead{
        \colhead{Line} & \colhead{Baselines} & \colhead{Peak R.A.}  & \colhead{Peak Dec.}  & \colhead{Maj, Min, P.A.} & \colhead{Inclination} \\ & & \colhead{$+5^\mathrm{h}35^\mathrm{m}21.5^\mathrm{s}$} & \colhead{$-5^\circ09'42\farcs0$} & \colhead{(au, au, deg)} & \colhead{(deg)}
    }
\startdata
    CO (3--2)      & $>70k\lambda$ & $1.19\pm0.01$ & $3.18\pm0.03$ & $790\pm30, 260\pm10, -8\pm1$ & $70.78\pm0.02$ \\
    HCN (4--3)     & All           & $1.15\pm0.02$ & $3.08\pm0.04$ & $850\pm40, 230\pm20, -4\pm2$ & $74.30\pm0.03$\\
    HCO$^+$ (4--3) & All           & $1.14\pm0.01$ & $3.19\pm0.02$ & $520\pm20, 240\pm10, -5\pm2$ & $62.51\pm0.03$\\
    HCO$^+$ (4--3) & $>70k\lambda$ & $1.14\pm0.01$ & $3.17\pm0.02$ & $540\pm20, 200\pm20, -3\pm2$ & $68.26\pm0.04$
\enddata
\tablecomments{Peak position is given as offset in arcsec from primary beam center. Maj, Min, and P.A. refer to the (FWHM of the) major and minor axis and position angle of the elliptical best-fit Gaussian. Position angle is defined as the rotation East of North of the blue-shifted side of the disk major axis. No results are given for CO (3--2) data containing all baselines, as the data were too contaminated to give a meaningful result, or CS (7--6) data, as the line was only detected at the $3\,\sigma$ level and showed no structure. }
\end{deluxetable*}

Assuming optically thin emission (consistent with the best-fit model presented in Section~\ref{sec:gasmod}) and Local Thermodynamic Equilibrium (LTE), the line-emitting gas mass, $M_\mathrm{gas}$, is given by 
\begin{equation}\label{eq:f2m}
    M_\mathrm{gas}=\frac{4\pi}{h\nu_0}\frac{F m d^2}{A_{u\ell} X_u},
\end{equation}
where $F$ is the integrated flux in the line (see Table \ref{tab:lineF}), $m$ is the mass of the emitting gas molecule, $d$ is the distance to the source, $h$ is the Planck constant, $\nu_0$ is the rest frequency of the line, $A_{u\ell}$ is the Einstein coefficient for the ($u$--$\ell$) transition and 
\begin{equation}\label{eq:Xu}
    X_u=\frac{N_u}{N_\mathrm{tot}}=(2J_u+1)\frac{\exp[-B_0J_u(J_u+1)hc/kT_\mathrm{ex}]}{kT_\mathrm{ex}/hcB_0}.
\end{equation}
In Equation \ref{eq:Xu}, $N_u$ is the number of molecules in the upper state, $N_\mathrm{tot}$ is the total number of molecules, $J_u$ is the quantum number of the upper level, $B_0$ is the rotational constant (in units of wavenumber), $h$ is the Planck constant, $c$ is the speed of light, $k$ is the Boltzmann constant, and $T_\mathrm{ex}$ is the excitation temperature. Values for $A_{u\ell}$ and $B_0$ were taken from molecular data made available by \citet{moldat}. The mass of the gas species in question must be scaled by the fractional abundance to obtain a total gas mass. The calculated mass is therefore dependent on the assumed fractional abundance of the gas species. Using the HCO$^+$ line, which has the highest signal-to-noise ratio, and a distance of 414\,pc we calculate a total disk mass of $0.013\pm0.001$ $M_\odot$ ($14\pm1\,M_\mathrm{Jupiter}$). This calculation uses an excitation temperature of 17\,K, which is consistent with the expected midplane temperature at the radius corresponding to the angular resolution (100\,au) and is consistent with the brightness temperature of the CO absorption of $\sim20$\,K cooler than the $\sim35$\,K background of the Integral Shaped Filament. The absorption seen in CO (3--2) indicates that the disk is slightly cooler than the background and that it is in front of the Integral Shaped Filament. The mass calculation also assumes an HCO$^+$ fractional abundance of $10^{-9}$. Due to large uncertainties in the abundance of HCO$^+$ relative to H$_2$, complicated by uncertainties in the vertical temperature structure of the disk (see Section \ref{sec:bftemp}), the systematic uncertainty in this measurement is likely much larger than the quoted statistical uncertainty. If the line is optically thick rather than thin as in our best-fit model, then this mass measurement should be considered a lower limit.

\begin{figure}
\begin{center}
    \includegraphics[width=\columnwidth]{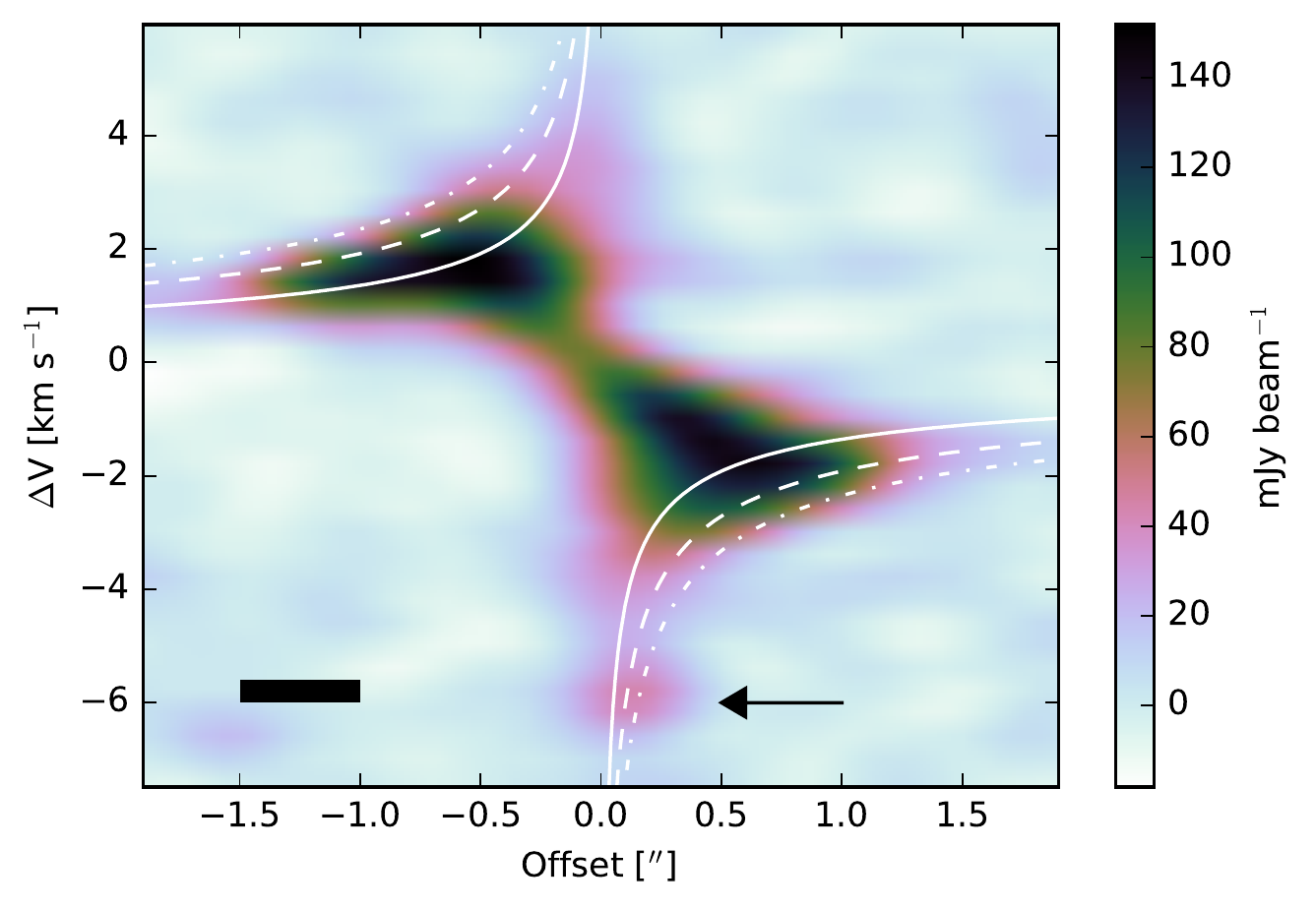}
    \caption{Position-velocity diagram along the disk major axis (shown in Figure \ref{fig:HCOpmoms}) of the HCO$^+$ (4--3) emission. Solid, dashed, and dot-dashed curves are Keplerian velocity profiles for stellar masses of 1\,$M_\odot$, 2\,$M_\odot$, and 3\,$M_\odot$, respectively, at an inclination of 68$^\circ$ and a systemic velocity of 10.67\,km\,s$^{-1}$ LSRK. The spatial and velocity resolution of 0\farcs5 and 0.41\,km\,s$^{-1}$ are indicated by the black box in the lower left corner. The black arrow indicates the position of the high velocity asymmetry.}
    \label{fig:PV}
\end{center}
\end{figure}

A position-velocity (P-V) diagram for HCO$^+$ (4--3) is shown in Figure \ref{fig:PV}. This diagram shows the position, as a function of velocity, of emission from a cut along the major axis of the disk. A clear asymmetry in the spectral extent of the lines can be seen with the blue-shifted side showing $3\,\sigma$ emission 3.8\,km\,s$^{-1}$ and 2.5\,km\,s$^{-1}$ further from systemic velocity than the red-shifted side in CO and HCO$^+$, respectively. This feature is particularly evident toward the bottom of the P-V diagram, indicated by the arrow in Figure \ref{fig:PV}. There is a distinct unresolved $47\pm6$\,mJy\,beam$^{-1}$ peak at a location $0\farcs11\pm0\farcs03$ to the north-west of the star and at a velocity around $-6$\,km\,s$^{-1}$ relative to the systemic velocity. We discuss possibilities for the nature of the feature in Section \ref{sec:hvf}. Overall, the P-V diagram suggests gas moving with a Keplerian velocity profile around a central star slightly more massive than 2\,M$_\odot$. This is significantly more massive than the mass inferred from its spectral type ($\sim1\,M_\odot$ for a K5 at 1--2\,Myr). There is some asymmetry in the shape of the red/blue-shifted sides of the line near systemic velocity. These channels show significant cloud contamination in CO (3--2), with the disk seen in absorption. This contamination may also be present in HCO$^+$, though below the noise threshold, and may be contributing to this asymmetry. 

\begin{figure}
\begin{center}
    \includegraphics[width=\columnwidth]{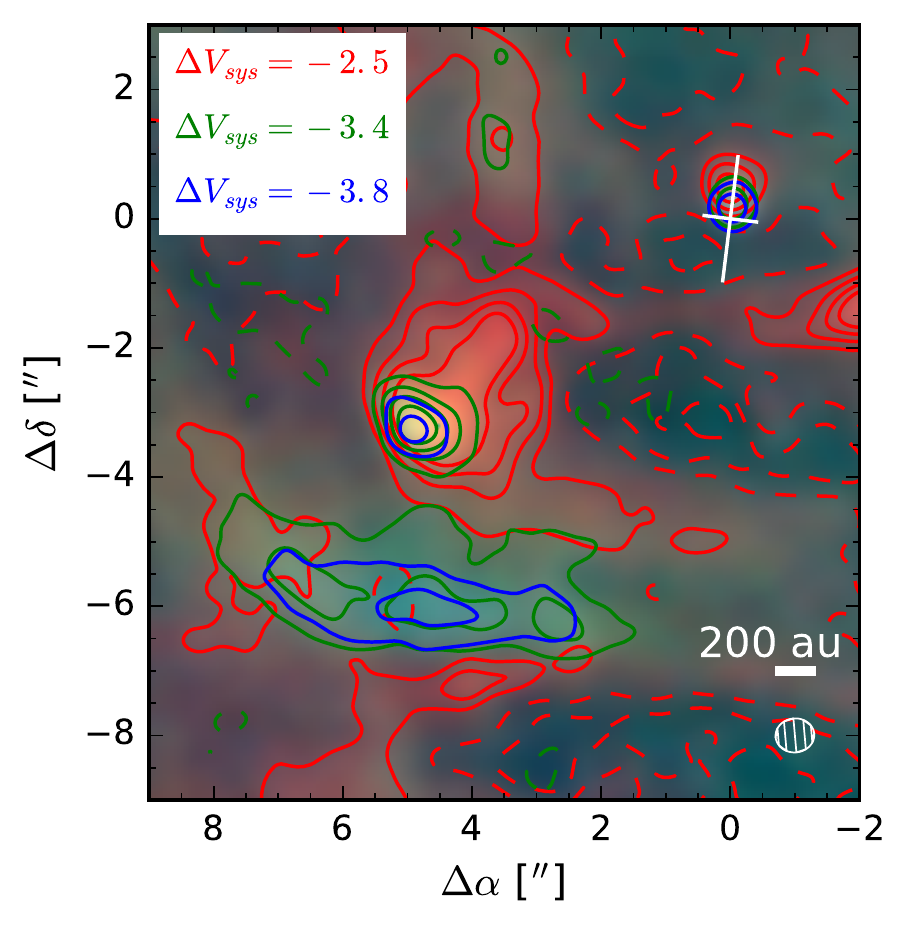}
    \caption{RGB image and contour map of CO (3--2) emission showing the detection of a possible blue-shifted outflow and associated bow shock. The location and approximate size, inclination, and position angle of the disk is shown by the white cross, while the outflow feature is located $\sim6''$ to the south-east. The velocity of the red, green, and blue channels/contours are given in the upper left corner in km\,s$^{-1}$ relative to the systemic velocity (Note: all three channels are blue-shifted relative to the systemic velocity). Contours are at $\pm10$, 20, 30 and 40\,$\sigma$ where $\sigma=5.5$\,mJy\,beam$^{-1}$. Dashed contours are negative. The FWHM size of the synthesized beam is shown in the bottom right corner along with a 200\,au scale bar.}
\label{fig:out}
\end{center}
\end{figure}

We also present the CO (3--2) detection of a possible outflow feature $\sim6''$ south-east of the disk, shown in Figure \ref{fig:out}. The feature is not detected in any of the other observed lines. With respect to the velocity of the main blue-shifted outflow lobe, a red-shifted tail can be seen pointing back toward the disk. There is also a blue shifted bow shape to the south-east of the feature, suggestive of a shock. If this feature is confirmed to be an outflow, it would illustrate the extreme youth of the system. Also, its orientation with respect to the disk is $\sim50^\circ$ away from the rotation axis. This could indicate complicated dynamics in the inner disk possibly caused by a companion. While the investigation of the feature is beyond the scope of this work, we present the observations here so that others may study it further.

\section{Modeling}
\label{mod}

Spatially and spectrally resolved observations of gas emission allow us to determine basic characteristics of the disk-star system. The line brightness as a function of position and velocity reflect the temperature and density structure of the gas disk, which we investigate through comparison with radiative transfer models of the disk structure. The spatially and spectrally resolved emission also encodes information about the mass of the central star and the inclination of the disk to the line of sight. In Section~\ref{sec:gasmod} we describe our physical model of the gas disk, and in Section~\ref{sec:fitproc} we describe the procedure used to fit the parameters to the data. 

\subsection{Gas Disk Model}\label{sec:gasmod}

We adopt the parameterization of disk temperature structure from \citet{temp} which was derived as an analytical approximation to the output of 2-D radiative transfer calculations for an optically thick dust disk. The global temperature profile is 
\begin{equation}\label{eq:Tstruct}
T_\mathrm{gas}(r,z)=\left\{
	\begin{array}{lr}
	T_a+(T_m-T_a)\left[\cos\frac{\pi z}{2z_q}\right]^{2\delta} & \mathrm{if} z < z_q\\
	T_a & \mathrm{if} z \geqslant z_q
	\end{array}
	\right.
\end{equation}
where the atmospheric temperature, $T_a$, is given by $T_a=T_{atm,150}(r/150\,\mathrm{au})^{q}$ and the mid-plane temperature, $T_m$, is given by $T_m=T_{mid,150}(r/150\,\mathrm{au})^{q}$. $T_{atm,150}$ and $T_{mid,150}$ are the temperatures at 150\,au and a height of $z=z_q$ and 0, respectively, and $q$ is a power law index. The height of the disk, controlled by $z_q$, is assumed to have a radial distribution described by a power law, $z_q=z_{q,150}(r/150\,\mathrm{au})^{1.3}$. Here we set $\delta$ to 1, although values $\delta\approx1-2$ have been used elsewhere. 

The gas surface density profile is assumed to be a power law with an exponential taper:
\begin{equation}
    \Sigma_\mathrm{gas}(r)=\frac{M_\mathrm{gas}(2-\gamma)}{2\pi R_\mathrm{c}^2}\left(\frac{r}{R_\mathrm{c}}\right)^{-\gamma}\exp\left[-\left(\frac{r}{R_\mathrm{c}}\right)^{2-\gamma}\right]
\end{equation}
where $R_\mathrm{c}$ sets the radial size of the gas disk, $\gamma$ is a power law index, and $M_\mathrm{gas}$ is the total gas mass. This tapered power law structure is consistent with theoretical predictions for the functional form of the density profile of an evolving viscous accretion disk \citep{density1,density2}, and is more physically plausible for an isolated disk than a truncated power law (i.e. a strict outer radius). 

We assume that $M_\mathrm{gas}=M_\mathrm{disk}$ as the dust makes up $\sim1\%$ of the disk mass. We also assume a molecular gas composition that is 80\% molecular hydrogen by weight with a mean molecular mass of 2.37. The density structure $\rho(r,z)$ is then calculated by solving the equation of hydrostatic equilibrium using the surface density profile and the temperature structure. In the mid-plane of the disk the temperatures are cold enough for gas to freeze onto dust grains. When modeling CO emission, we mimic this freeze out by multiplying the density by a factor of $10^{-5}$ wherever the temperature is less than 19\,K \citep[thus lowering the relative abundance of CO to $10^{-9}$, consistent with][]{Wal10}. The upper surface is governed by photodissociation by stellar radiation. We thus drop the density similarly wherever the vertical hydrogen column density from the surface of the disk is less than $1.3\times10^{21}\,\mathrm{cm}^{-2}$. The freeze-out temperature and column density values were adopted from an analysis of the HD 163296 disk by \cite{qi11}, while other values have been used elsewhere \citep[e.g. $5\times10^{20}\,\mathrm{cm}^{-2}$ and $9\times10^{20}\,\mathrm{cm}^{-2}$ for a vertically isothermal and structured disk, respectively; see][]{IDL3}. We use the same freeze-out and photodissociation values for CO and HCO$^+$ as CO is a chemical precursor of HCO$^+$ \citep[$\mathrm{H}_3^++\mathrm{CO}\rightarrow\mathrm{HCO}^++\mathrm{H}_2$,][]{Cle15,Yu16}. For HCN, we apply a 60\,K freeze-out temperature \citep{van14,Gar06} and a $9.5\times10^{21}\,\mathrm{cm}^{-2}$ photodissociation column density \citep{Fuente1993}. We also applied a lower column density threshold of $1\times10^{23}\,\mathrm{cm}^{-2}$ (L.~I. Cleeves, private communication) to mimic HCN photodesorption such that even if the temperature is below the HCN freeze-out threshold, if the vertical column density is less than the photodesorption threshold, freeze-out is not applied.

We then calculate the Keplerian velocity profile with a small correction for pressure and height (while still restricting orbits to be parallel to the plane of the disk). Since a gradient in pressure results in a force we must correct for this force according to 
\begin{equation}\label{eq:kepres}
    \frac{v_\phi^2}{r}=\frac{GM_*r}{(r^2+z^2)^{3/2}}+\frac{1}{\rho_\mathrm{gas}}\frac{\partial P_\mathrm{gas}}{\partial r}.
\end{equation}
For typical conditions in the outer disk (30\,K at $\sim500$\,au) using a simple power law density profile, this term reduces the velocity of the gas by $\sim1\%$ \citep{Arm09}. The exponential tail used in this model enhances this effect in the outer disk \citep{IDL3}. We assume that the gas kinematics are described by this modified Keplerian velocity field with no component of the velocity in the vertical direction. This assumption is valid as long as $M_{disk}\ll M_*$ (i.e. the disk is not self gravitating) which is consistent with constraints from continuum observations. 

At this point, the physical structure (temperature, density and velocity) of the disk is completely determined. To translate the physical structure (temperature, density, and velocity) into a sky-projected image, we then use a ray tracing code originally written in IDL by \citet{IDL2,IDL3} and translated into Python by \citet{weakTurb}. It assumes LTE, which is not always valid in protoplanetary disks, although \cite{LTE} showed that it is appropriate for CO. We investigate the robustness of the LTE assumption for the HCO$^+$ (4--3) and HCN (4--3) line in Section \ref{sec:LIME}. The computational efficiency of the LTE assumption enables the use of the Markov Chain Monte Carlo (MCMC) technique which can be used to characterize the posterior distribution on each parameter, taking into account prior knowledge and degeneracies between parameters. 

We include a Doppler shift along the spectral dimension to account for the systemic velocity, $v_{sys}$, and scale, shift, and rotate the image to account for the distance to the source $d$, position offset from the center of the image $\Delta\alpha$ and $\Delta\delta$, and position angle $PA$. The free parameters of our model are summarized in Table~\ref{tab:param}. The spatial resolution of the model is set to $\sim1/10$ the size of the synthesized beam. We Hanning smooth the model image and simulate observations using the MIRIAD task \texttt{uvmodel}. We then calculate a $\chi^2$ statistic by comparing the data and sampled model in the visibility domain. 

\begin{deluxetable*}{lcllcl}
    \tablewidth{0pt}
    \tablecaption{Model Parameters and Priors\label{tab:param}}
    \tablehead{\colhead{Symbol} & \colhead{Parameter} & \colhead{Prior} & \colhead{Symbol} & \colhead{Parameter} & \colhead{Prior} } 
\startdata
$X_\mathrm{mol}$ & relative abundance & $\log$-uniform or fixed\tablenotemark{a}& $R_\mathrm{c}/R_\mathrm{mol}$ & critical/outer molecule density radius & $\log$-uniform \\
$M_\mathrm{disk}$ & disk gas mass & $\log$-normal\tablenotemark{b} or fixed\tablenotemark{a} & $\gamma$ & radial density power law index & fixed\tablenotemark{a}\\
$M_\mathrm{*}$ & mass of star & $\log$-uniform & $q$ & radial temperature power law index & uniform \\
$v_\mathrm{sys}$ & systemic velocity & none\tablenotemark{c} & $z_{q,150}$ & disk height at 150 au & fixed\tablenotemark{a}\\
$v_{turb}$ & turbulence velocity & fixed\tablenotemark{a} & $T_\mathrm{mid,150}$ & mid-plane temperature at 150 au & fixed\tablenotemark{a}\\
$d$ & distance & fixed\tablenotemark{a} & $T_\mathrm{atm,150}$ & atmospheric temperature at 150 au & uniform \\
$i$ & inclination & uniform & $\Delta\alpha$ & disk offset in RA from center of image & none\tablenotemark{c}\\
$PA$ & position angle & uniform & $\Delta\delta$ & disk offset in Dec from center of image & none\tablenotemark{c}
\enddata
\tablenotetext{a}{These parameters were fixed based on prior knowledge.}
\tablenotetext{b}{Log-normal prior on the disk mass was centered on previous continuum mass measurements $\mu= 0.0445\,\mathrm{M_\odot}$ \citep{MassiveOrion,ALMAOri}, $\sigma = 1 \mathrm{dex}$.}
\tablenotetext{c}{These parameters were fit with a grid search as described in Section \ref{sec:fitproc} and were fixed during the MCMC fit.}
\end{deluxetable*}

\subsection{Fitting Procedure}
\label{sec:fitproc}

We first performed a simple grid search for the x-y position offset from the center of the image and the systemic velocity, using a nominal set of temperature and density parameters within the range of values typically observed for protoplanetary disks, along with an inclination and position angle determined from the images and previous observations. We used only the HCO$^+$ (4--3) line for these preliminary fits, since that line has a higher signal-to-noise ratio than the HCN (4--3) line and significantly less contamination than the CO (3--2) line. We took care to exclude channels with the high velocity excess emission discussed in Section \ref{res} when performing all fits. Hence, channels with velocities less than or equal to $-4.3$\,km\,s$^{-1}$ (relative to systemic velocity) were excluded from fits to the HCO$^+$ (4--3) data. These fits yielded a systemic velocity of $10.7\pm0.1$\,km\,s$^{-1}$ and a $\Delta\alpha$ and $\Delta\delta$ of $1\farcs17\pm0\farcs02$ and $3\farcs13\pm0\farcs02$, respectively (the J2000 pointing center of the observations is $\alpha$ = $5^\mathrm{h}35^\mathrm{m}21.5^\mathrm{s}$, $\delta$ = $-5^\circ09'42\farcs0$). The grid searches for systemic velocity and position offset had step sizes of 0.01 km\,s$^{-1}$ and 0\farcs01, respectively. 

We then fit the model of the disk temperature, density, and velocity to the data using a Markov Chain Monte Carlo (MCMC) algorithm with an affine invariant sampler. Priors associated with each parameter are given in Table~\ref{tab:param}. We exclude the first $\gtrsim3$ autocorrelation times for burn-in and use at least 3 more in our statistical analysis. In some cases, individual walkers that had not burned in even after $\gtrsim3$ autocorrelation times had to be removed. We used the open source \texttt{emcee} software package, written by \citet{emcee}. This package is based on the affine invariant sampler described by \citet{afmcmc} which probes degenerate parameter space more efficiently than the classical Metropolis-Hastings algorithm with a Gibbs sampler. Figure \ref{fig:HCOptri} shows an example of a triangle (or corner) plot showing the 1- and 2-D posterior distributions made with the associated open source python package \texttt{corner.py} \citep{corner}. 

Since we modeled the three molecular lines separately, it was not possible to meaningfully constrain the vertical temperature structure of the disk. We therefore fixed $T_\mathrm{mid,150}$ and $z_{q,150}$ at 17.5 K and 70 au respectively, which are the best-fit values found in an analysis of a much closer and therefore higher-quality set of observations of a protoplanetary disk around a star with very similar mass \citep[$2.3\,M_\odot$,][]{weakTurb}. Our observations do not have high enough spectral resolution to constrain the turbulent linewidth so we fixed $v_{turb}$ at 1\% of the sound speed. For optically thin lines the power law indices describing the dependence of temperature and surface density with radius are degenerate. Hence, we fixed $\gamma$ at $1$, a typical value for disks in Ophiuchus \citep{OphStruct,OphStruct2}. In order to have consistency in the disk structure between the different lines, we fixed $R_\mathrm{c}$ at 600\,au, the outer radius of the scattered light disk \citep[][also comparable to the radius of the 3\,$\sigma$ contour in the moment 0 maps]{HST}, and instead fit $R_\mathrm{mol}$, a sharp outer radius for each molecule. We attempted to fit all three lines simultaneously but were unable to identify a model with insignificant residuals. While fitting CO and HCO$^+$ together could produce a good fit, the addition of HCN greatly increased the residuals. This is an indication that a more complicated treatment of the photochemistry of HCN may be needed to optimally reproduce the data though this is beyond the scope of the current work. 

Baselines shorter than $70\,k\lambda$ were excluded from CO (3--2) and HCO$^+$ (4--3) observations to minimize cloud contamination while all baselines were used for the HCN (4--3) observations (see Section \ref{res} for a discussion of contamination). While the removal of baselines shorter than 70\,k$\lambda$ reduces the total flux by a factor of $\sim2$ and causes increased spatial filtering that may miss some extended flux, it is necessary to avoid the cloud contamination since the contribution of the two components to the short-baseline flux cannot be disentangled in the visibility domain. A test that involved fitting HCN with and without baselines shorter than 70\,k$\lambda$ resulted in outer radii, and all other parameters, that were consistent to within $<1\sigma$. We use the optically thick CO (3--2) line to estimate the temperature structure and minimum disk mass, and the optically thin HCO$^+$ (4--3) and HCN (4--3) lines to estimate their respective fractional abundances. All three of the lines allow us to characterize the mass of the central star and the radius, inclination, and position angle of the disk.

\subsubsection{HCO$^+$ (4--3) Fit}\label{sec:hcopfit}

Since the abundance of HCO$^+$ is less well constrained than CO, we allowed $X_\mathrm{HCO+}$ to vary while fixing $M_\mathrm{disk}$ at the value of 0.0445\,M$_\sun$ consistent with the continuum measurements. This approach allowed us to determine the accuracy of the literature values for the relative abundance of HCO$^+$ with the caveat that continuum mass measurements are inherently lower limits due to uncertainties in the gas to dust ratio, potentially optically thick emission, and large bodies which do not emit at millimeter wavelengths. We also assume a constant abundance throughout the disk (in between the photodissociation and freeze-out surfaces) which is a simplification of the actual vertical and radial structure, as discussed in more detail in Section \ref{sec:bftemp}.

\begin{figure*}
    \begin{center}
    \includegraphics[width=\textwidth]{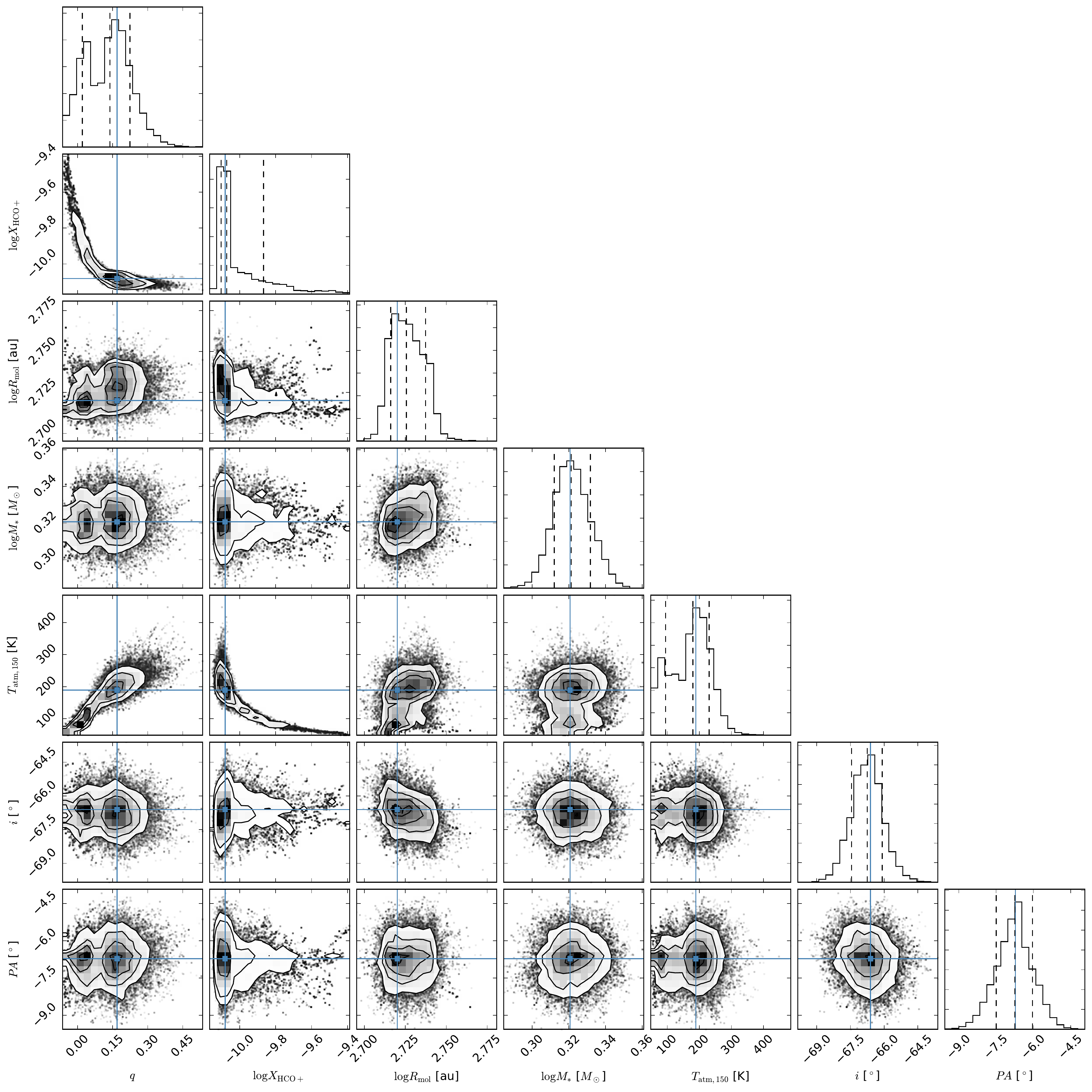}
    \caption{Triangle plot showing the posterior probability distributions of the seven parameter fit to the HCO$^+$ (4--3) emission (of which results are given in Table~\ref{tab:bfp}). Plots along the diagonal show the posterior distribution of each parameter individually while plots below the diagonal compare a pair of parameters showing potential degeneracies. Solid blue lines show the best fit parameter values while dashed lines show the median and $\pm1\,\sigma$ values.}
    \label{fig:HCOptri}
    \end{center}
\end{figure*}

\begin{figure*}
\begin{center}
\includegraphics[width=\textwidth]{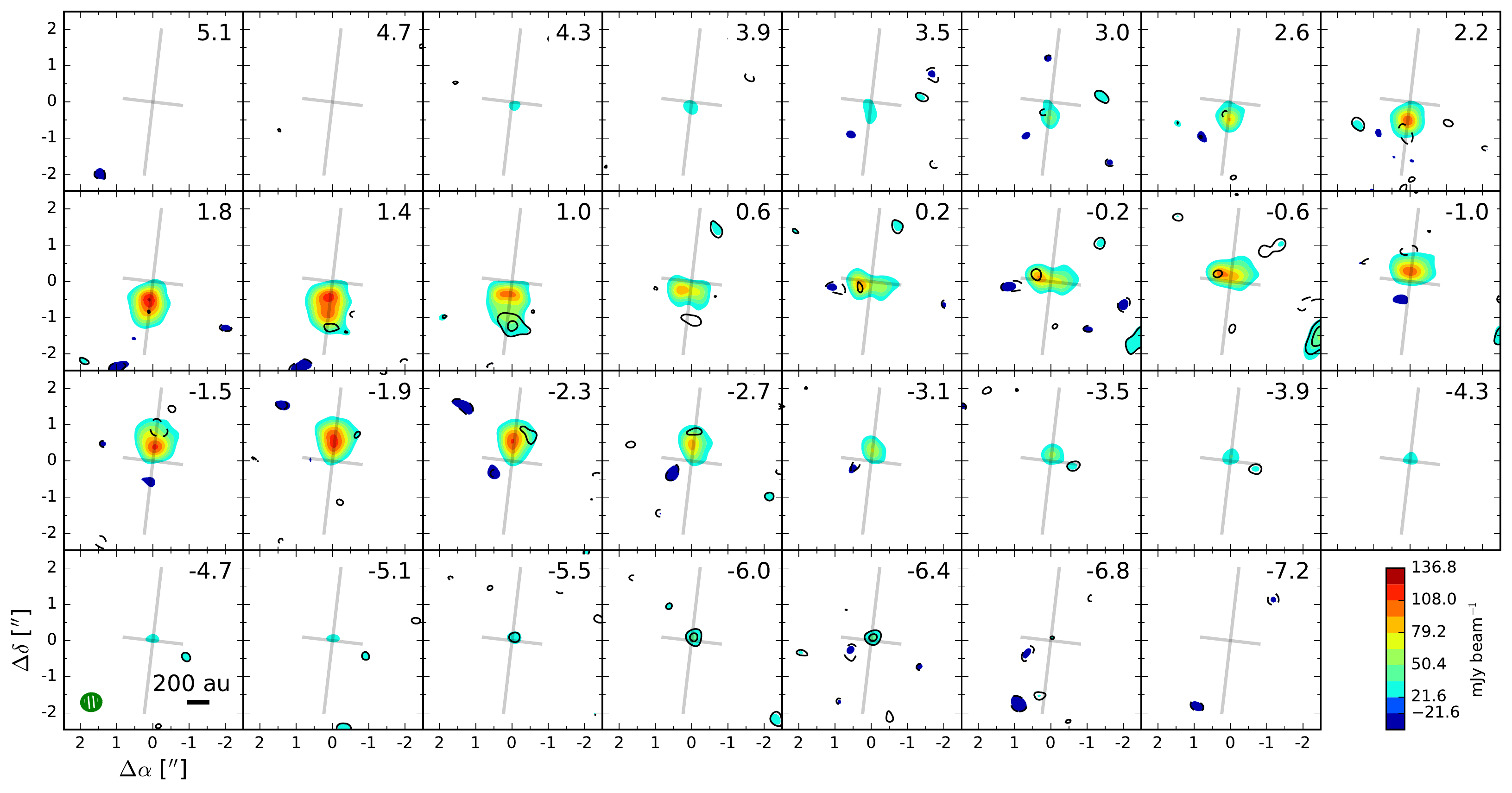}
\caption{Naturally weighted channel maps of HCO$^+$ (4--3) emission excluding baselines shorter than $70\,k\lambda$. Colors and contours start at $\pm3\,\sigma$ with increments of $2\,\sigma$ up to $17\,\sigma$ where $1\,\sigma$ is 7.2\,mJy\,beam$^{-1}$. Dashed contours correspond to negative values. Colors show the data while contours are residuals from the best fit, given in Table~\ref{tab:bfp}. $\Delta V$ from systemic velocity (+10.7\,km\,s$^{-1}$ LSRK) is given in the upper right corner. Channels with $\Delta V\leq-4.3$\,km\,s$^{-1}$ were excluded when fitting. The FWHM size of the synthesized beam is shown in the bottom left corner along with a 200 au scale bar. The grey cross is centered on the star position and shows the position angle of the disk and the major-to-minor axis ratio.}
\label{fig:HCOpres}
\end{center}
\end{figure*}

As seen in Figure \ref{fig:HCOptri} and Table~\ref{tab:bfp}, all parameters are well constrained with noticeable degeneracies between parameters affecting the temperature structure ($q$ and $T_\mathrm{atm,150}$) and density structure ($X_\mathrm{HCO+}$ and $R_\mathrm{mol}$). There is no degeneracy between inclination, $i$, and stellar mass, $M_\mathrm{*}$, due to the high spatial resolution of the data which allows the disk ellipticity (which is assumed to be due to the inclination) to be precisely measured. Channel maps showing residuals of HCO$^+$ (4--3) from the best fit model are shown in Figure~\ref{fig:HCOpres}. Significant residuals in high velocity blue-shifted channels are expected as we did not model the high-velocity ``clump" of emission located in those channels. Elsewhere, $3\,\sigma$ residuals are located only in channels with significant cloud contamination in CO. There may be some associated contamination in HCO$^+$ which could be causing asymmetries in the disk emission in those channels. 

\subsubsection{CO (3--2) Fit}\label{sec:cofit}

Even after excluding baselines shorter than $70k\,\lambda$, CO (3--2) emission was still highly contaminated in channels near the systemic velocity. In addition, fits to the individual CO (3--2) line, only modeling uncontaminated channels, showed an asymmetry between the red- and blue-shifted sides of the disk in a similar manner to the asymmetry seen due to the high-velocity feature in the HCO$^+$ (4--3) emission. In CO (3--2), however, this excess emission is spread across many channels on the blue-shifted side of the line rather than as an isolated feature in the HCO$^+$ (4--3) line, a difference likely due to optical depth broadening in the much more optically thick CO (3--2) tracer. Thus, when fitting CO (3--2), we excluded the entire blue-shifted side of the line as well as the contaminated channels near the systemic velocity. We therefore only model channels with velocities $\geq$+3.4\,km\,s$^{-1}$ relative to the systemic velocity. 

We assumed a fractional abundance $X_\mathrm{CO}$ of $10^{-4}$, a typical value for the ISM \citep{abund,abund2}. We also placed a log-normal prior on the disk mass centered on the previously measured mass of 0.0445 $M_\odot$ \citep[measured from continuum observations]{MassiveOrion,ALMAOri} with a standard deviation of 1 dex. This approach was done to constrain the disk mass near the previously measured value but still allow the parameter to vary.

\begin{figure*}
    \begin{center}
    \includegraphics[width=\textwidth]{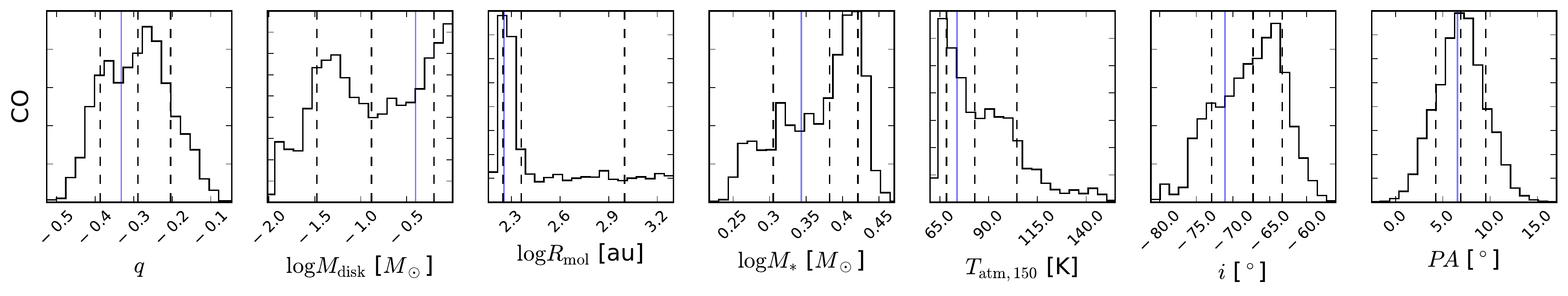}
    \includegraphics[width=\textwidth]{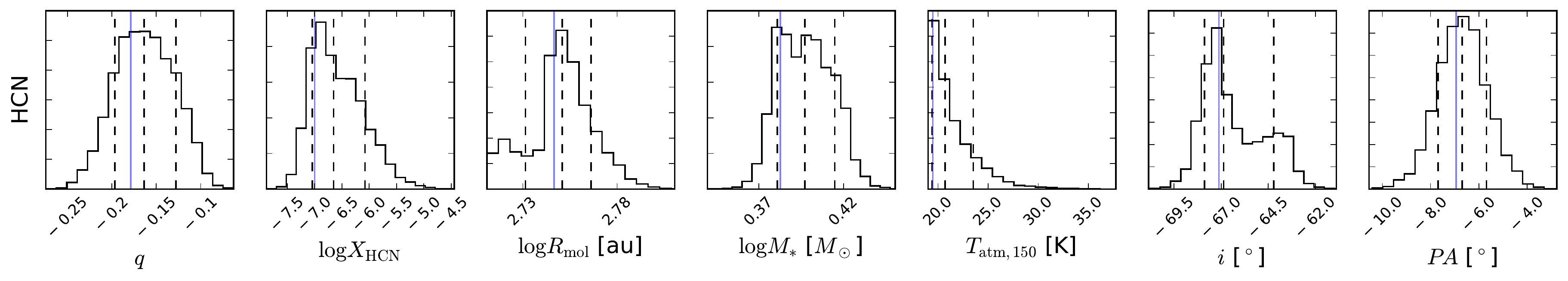}
    \caption{Histograms showing the posterior probability distributions of the 7 parameter fits to the CO (3--2) (top) and HCN (4--3) (bottom) emission, respectively. Solid blue lines indicate the best fit parameter, while dashed black lines indicate the median and $\pm1\sigma$ values (given in Table~\ref{tab:bfp}).}
    \label{fig:hist}
    \end{center}
\end{figure*}

Histograms showing the posterior distributions of the individual line fit are shown in the top row of Figure~\ref{fig:hist}. Best fit and median values with $1\,\sigma$ uncertainties are given in Table~\ref{tab:bfp}. Since CO (3--2) is optically thick, we would expect this line to provide only a lower limit on the mass which is clearly seen in the histogram as a sharp drop off in the posterior distribution near 0.02 $M_\odot$ ($\log M_\mathrm{disk}/M_\odot\sim-1.7$). Also, since the outer radius is constrained mainly by the channels near the systemic velocity, which are not included in our fit, $R_\mathrm{mol}$ is not well constrained. The inclination and position angle have much larger uncertainties than those of the fit to the HCO$^+$ (4--3) line for the same reason. The stellar mass is also less well constrained because we are only fitting channels which are marginally spatially resolved. In this fit a degeneracy between inclination, $i$, and stellar mass, $M_\mathrm{*}$, is seen due to the poorly constrained inclination. 

\begin{figure*}
\begin{center}
\includegraphics[width=\textwidth]{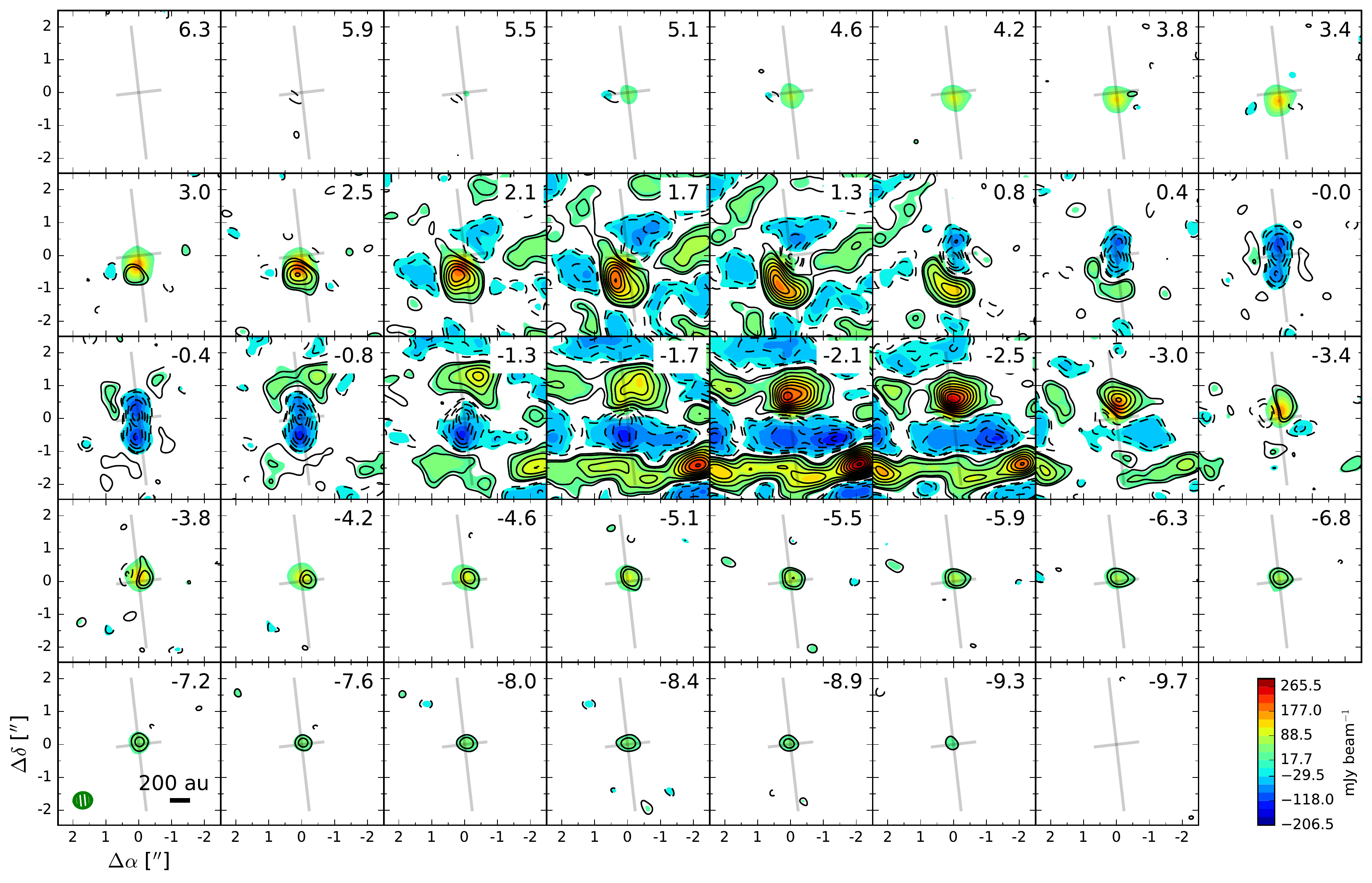}
\caption{Naturally weighted channel maps of CO (3--2) emission excluding baselines shorter than $70\,k\lambda$. Colors and contours are at 3, 5, 10, 15\ldots$50\,\sigma$ where $1\,\sigma$ is 5.9\,mJy\,beam$^{-1}$. Colors show the data while contours are residuals from the best fit to the CO (3--2) transition, given in Table~\ref{tab:bfp}. Symbols are the same as in Figure~\ref{fig:HCOpres}. Only channels with $\Delta V\geq3.4$\,km\,s$^{-1}$ were fitted.}
\label{fig:COres}
\end{center}
\end{figure*}

Channel maps showing residuals from the best fit model, compared to the corresponding data, are shown in Figure~\ref{fig:COres}. Significant residuals in high velocity blue-shifted channels and those near systemic velocity are expected as we did not model those channels. Residuals near the systemic velocity are due to cloud contamination while those in blue-shifted channels are due to the previously discussed asymmetry. Residuals in the channels which are modeled are minimal.

\subsubsection{HCN (4--3) Fit}\label{sec:HCNfit}

All baselines were included in the HCN (4--3) fit, as there is no significant cloud contamination in this line. Also, no excess high velocity blue-shifted emission was noted in this line so all channels were used in the fit. As with our fit to HCO$^+$ (4--3), we chose to fix the disk mass at the mass inferred from continuum measurements and fit the abundance. 

After preliminary fits, we noticed significant residuals that were antisymmetric about the systemic velocity, indicative of a velocity offset. We then re-fit the systemic velocity for this line and found a best fit value of 10.49\,km\,s$^{-1}$. This $\sim0.2$\,km\,s$^{-1}$ offset, roughly a third of a channel, is within the uncertainty on the systemic velocity derived from HCO$^+$, and is therefore likely to be a calibration or noise issue rather than physical in nature. 

\begin{figure*}[h!]
\begin{center}
    \includegraphics[width=\textwidth]{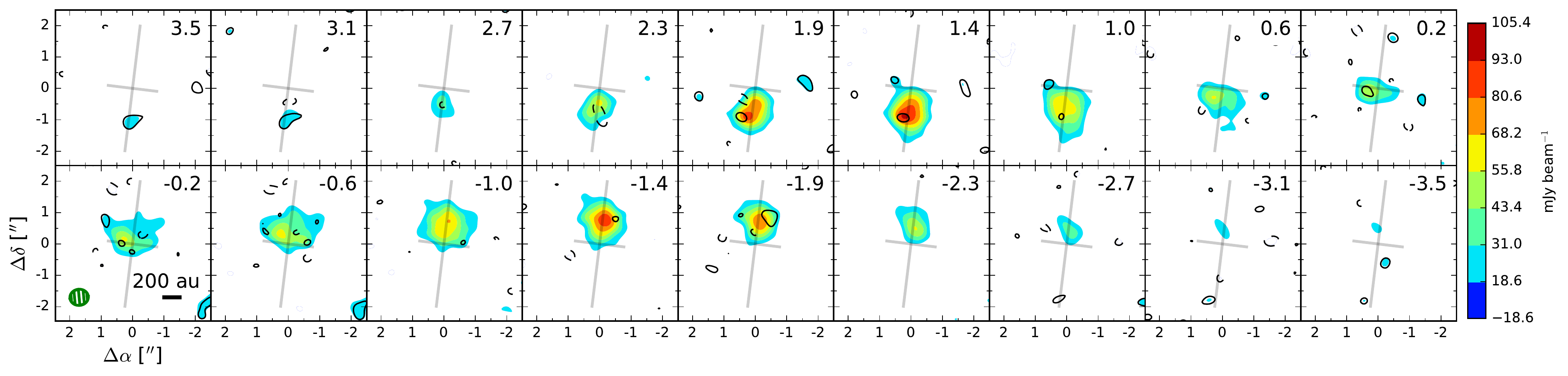}
    \caption{Naturally weighted channel maps of HCN (4--3). Colors and contours are at $\pm$3, 5, 7\ldots15\,$\sigma$ where $1\,\sigma$ is 6.2\,mJy\,beam$^{-1}$. Contours are residuals from the best fit to the HCN (4--3) transition, given in Table~\ref{tab:bfp}. Symbols are the same as in Figure~\ref{fig:HCOpres}.}
    \label{fig:HCNres}
\end{center}
\end{figure*}

Histograms showing the posterior distributions of the individual line fit are shown in the bottom row of Figure~\ref{fig:hist}. Best-fit and median values with $1\,\sigma$ uncertainties are given in Table~\ref{tab:bfp}. No significant degeneracies between parameters were noted. Channel maps showing residuals from the best fit model, compared to the corresponding data, are shown in Figure~\ref{fig:HCNres}. Small areas of $3\,\sigma$ residuals located in the outer southern limb of the disk are similar to those seen in HCO$^+$ (4--3) and the asymmetry seen in the continuum emission \citep[see Figure 3 of ][]{ALMAOri}. The best-fit model is optically thin. 

\begin{deluxetable*}{lcccccccc}[h!]
    \tablewidth{0pt}
    \tablecolumns{9}
    \tablecaption{Best Fit Parameters\label{tab:bfp}}
    \tablehead{\colhead{Line} & \colhead{$q$} & \colhead{$M_\mathrm{disk}$ $(\mathrm{M_\odot})$} & \colhead{$R_\mathrm{mol}$ $(\mathrm{au})$} & \colhead{$M_*$ $(\mathrm{M_\odot})$} & \colhead{$T_\mathrm{atm,150}$ $(\mathrm{K})$} & \colhead{$\log X_\mathrm{mol}$} & \colhead{$i$ $(\mathrm{deg})$} & \colhead{$PA$ $(\mathrm{deg})$} }
    \startdata
    HCO$^+$ & $0.14^{+0.08}_{-0.11}$ & $[0.0445]$ &$530\pm10$ &$2.10^{+0.05}_{-0.04}$ &$180_{-90}^{+50}$ &$-10.07_{-0.03}^{+0.20}$ &$-66.7\pm0.7$ &$-6.7\pm0.7$ \\
(best) & $0.17$ & $[0.0445]$ & $520$ & $2.09$ & $190$ & $-10.08$ & $-66.6$ & $-6.7$ \\
    CO &$-0.29_{-0.10}^{+0.09}$ &$0.13^{+0.50}_{-0.01}$ &$230^{+770}_{-50}$ &$2.4_{-0.4}^{+0.2}$ &$80^{+20}_{-10}$ &$[-4]$ &$-67_{-6}^{+4}$ &$7\pm3$ \\
(best) & $-0.33$ & $0.39$ & $180$ & $2.2$ & $70$ & $[-4]$ & $-71$ & $7$ \\
    HCN & $-0.16_{-0.03}^{+0.04}$ & $[0.0445]$ &$560\pm20$ &$2.50_{-0.09}^{+0.10}$ &$21_{-1}^{+3}$ &$-6.7_{-0.4}^{+0.6}$ & $-67_{-1}^{+3}$ & $-6.7\pm1.0$ \\
(best) & $-0.18$ & $[0.0445]$ & $560$ & $2.41$ & $19$ & $-6.7$ & $-67$ & $-7.0$ 
    \enddata
    \tablecomments{In all fits $d$ was fixed at 414\,pc, $R_\mathrm{c}$ at 600\,au, $T_\mathrm{mid,150}$ at 17.5\,K, $z_{q,150}$ at 70\,au, $v_{turb}$ at $0.01\,c_s$, $\gamma$ at 1, $v_{sys}$ at 10.67\,km\,s$^{-1}$ (for HCO$^+$ and CO) and 10.49\,km\,s$^{-1}$ (for HCN, see Section \ref{sec:HCNfit}), and $\Delta\alpha$ and $\Delta\delta$ at 1\farcs17 and 3\farcs13 respectively. Individual parameters in square brackets were also fixed. Stated uncertainties do not include the ALMA absolute flux uncertainty of $\sim10\%$ or the distance uncertainty of $1.7\%$. }
\end{deluxetable*}

\subsection{Evaluation of the LTE Assumption}\label{sec:LIME}

\begin{figure*}
\begin{center}
\includegraphics[width=0.49\textwidth]{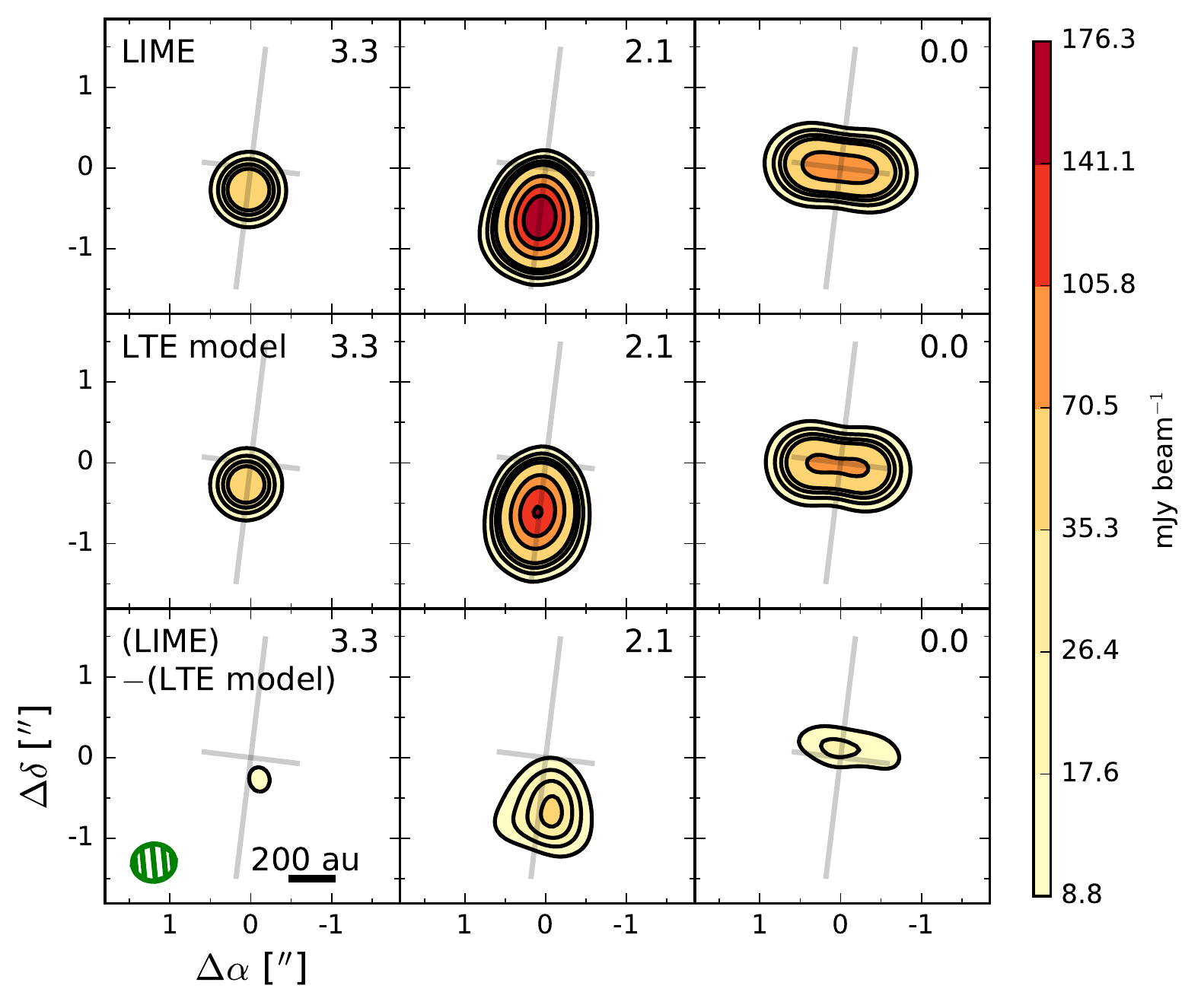}
\includegraphics[width=0.49\textwidth]{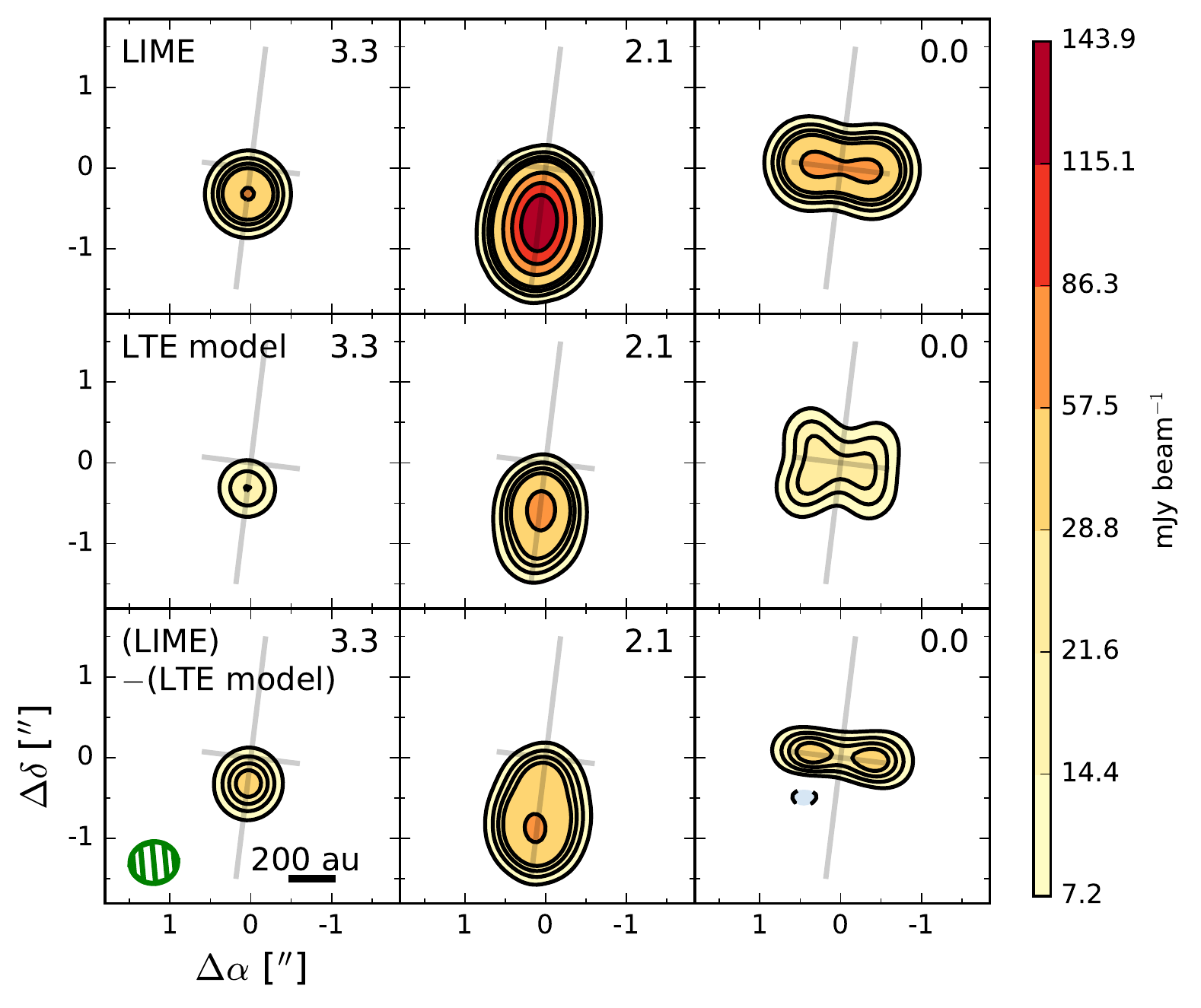}
\caption{Channel maps showing simulated observations of HCO$^+$ (4--3) (left) and HCN (4--3) (right) created using the best-fit parameters with LIME (top), our LTE model (middle), and the difference (bottom). The significance of the residuals demonstrate that LTE is a good assumption for HCO$^+$ (4--3) but not for HCN (4--3). Contours are at $\pm5$, 10, 15, 20, 40, 60, 80, and 100\% of the peak line flux (of the LIME model) where 5\% is 8.8 (9.7)\,mJy\,beam$^{-1}$ or equivalently 1.2 (1.5) times the rms noise in the HCO$^+$ (HCN) observations. $\Delta$V from the systemic velocity in km\,s$^{-1}$ is given in the upper right corner. The FWHM size of the synthesized beam is shown in the bottom left corner along with a 200 au scale bar. The gray cross is centered on the star position and shows the position angle of the disk and major-to-minor axis ratio. }
\label{fig:limecomp}
\end{center}
\end{figure*}

As discussed in Section \ref{sec:gasmod}, the assumption of local thermodynamic equilibrium (LTE) has been shown by \citet{LTE} to be appropriate for CO, but not for HCO$^+$ and HCN. We therefore investigated the differences between our LTE model and a non-LTE model using LIME \citep[LIne Modeling Engine,][]{LIME}. Figure~\ref{fig:limecomp} compares images of the best fit models to both the HCO$^+$ (4--3) and HCN (4--3) emission generated by the LTE model discussed in Section \ref{sec:gasmod} and LIME. After Hanning smoothing the full resolution images, observations were simulated using the MIRIAD task \texttt{uvmodel}. Three channel maps are shown: one at the systemic velocity (line-center), one near peak intensity, and one at the high-velocity line wing. 

For HCO$^+$ (4--3), the models differ by $\lesssim10\%$ of the line peak ($\lesssim2\,\sigma$ with respect to the noise threshold of our observations) in all but three channels near peak intensity on both sides of the line two of which show 15\% residuals and one shows 20\% residuals ($\sim3\,\sigma$ and $\sim5\,\sigma$, respectively). Overall, this test shows that our assumption of LTE should not drastically affect our results, and that the advantage in computational efficiency and the statistical characterization it allowed outweigh the small inaccuracies in our model.

On the other hand, for HCN (4--3) our LTE model is significantly fainter than the LIME model throughout the entire line. Two possible causes for this discrepancy are that the line is not in LTE, or that the way we implemented the abundance structure is not close enough to reality. Both factors likely play a role, and future modeling using a non-LTE radiative transfer code would be beneficial for better constraining the abundance of HCN in the disk. 

\section{Discussion}
\label{disc}

\subsection{Best Fit Temperature and Abundance Structure}\label{sec:bftemp}

Comparing the fits to the three molecules, we can see a few interesting trends in the temperature profiles (parameters are give in Table \ref{tab:bfp}). First, the value of $q$ in the fit to the HCO$^+$ emission was $\gtrsim0$. The expected value for $q$ in a geometrically flat, optically thin disk is $-0.5$ while measured values vary from $-0.6$ to $-0.3$ \citep{temp,IDL2,IDL} due to the flared structure of the disk and high optical depth. Our best-fit value when fitting the CO line is consistent with these values. The slightly positive value of $q$ derived for HCO$^+$ is most likely due to the effect of observing an optically thin line of a molecule that freezes out in the midplane. As \citet{sch16} explain in a similar investigation of the molecular line emission from the disk around TW Hya, a flat temperature profile is expected in the case where most of the detected emission originates from the layer of the disk just above the freeze-out temperature. Such observations effectively detect the surface snow line in the outer disk, which is at a constant temperature independent of radius. 

The best-fit atmospheric temperature for the CO line ($80^{+20}_{-10}$\,K), while consistent with other massive disks \citep[e.g.][]{weakTurb,IDL3}, is significantly cooler than the atmospheric temperature of our fit to HCO$^+$ ($180_{-90}^{+50}$\,K). The vertical temperature profile derived from CO is likely more reliable, since the low optical depth of the HCO$^+$ (4--3) line introduces a degeneracy between temperature and density (visible in Figure \ref{fig:HCOptri}). A simultaneous fit to all three spectral lines would provide a more robust constraint on the vertical temperature gradient in the disk. 

Since the best-fit models of HCO$^+$ and HCN are optically thin, we can also investigate the abundances of these molecules based on our models of the molecular emission. As outlined in Section \ref{sec:gasmod}, we assume a constant fractional abundance throughout the disk for all gas species (modified by prescriptions for freeze-out, photodissociation, and photodesorption effects). While this is a reasonable assumption for CO, the chemistry of HCO$^+$ and HCN causes their structure to be much more complicated. \citet{HCOpCle} and \citet{ionHCN} have both simulated the chemistry in protoplanetary disks with different levels of ionization (either from external sources or the central star), incorporating a steady-state chemical network. Both studies produce a vertical HCO$^+$ fractional abundance profile which, with increasing height, increases by several orders of magnitude, then decreases again. In these studies, the fractional abundances relative to H$_2$ peak at $10^{-9}$ and $10^{-6}$, respectively, which are somewhat higher than our measured HCO$^+$ abundance of $10^{-10.07_{-0.03}^{+0.20}}$. This may reflect our constant abundance structure effectively averaging over the more complicated vertical structure of \citet{HCOpCle} and \citet{ionHCN}.

While the vertical CO abundance in the models does follow a similar profile due to freeze-out in the midplane and photoionization near the disk surface, the gradient is much more exaggerated in HCO$^+$ due to the chemical reactions and ionization which produce the species. The profile also narrows (with respect to height) with increasing levels of ionization. In simulations by \citet{HCOpCle}, the HCO$^+$ fractional abundance even showed a two-layered structure with the main peak coincident with the peak of the CO progenitor gas and a weaker secondary peak near the surface of the disk where the ionization rate was higher. Since the vertical structure of this disk is not spatially resolved, we cannot distinguish between a one- and two-layer HCO$^+$ abundance structure with existing observations. Higher-resolution observations, however, might be able to confirm this feature of the chemical modeling. 

The vertical abundance structure of HCN is likely more complicated than our assumed constant abundance with photodissociation and photodesorption/freeze-out. We experimented with implementing an HCN abundance structure similar to \citet{Wal10} with a simple drop in abundance below the $z/r=0.25$ surface. This unfortunately caused an extreme deficit of emission in the outer disk due to the photodissociation surface falling below the $z/r=0.25$ surface in the inner $\sim$100\,au while the HCN emission stretches much farther out in the disk. A variable temperature structure with our more physical photodesorption and photodissociation thresholds, described in Section \ref{sec:gasmod}, was able to produce a better fit to the data, but only with a physically unrealistic temperature structure with an atmospheric temperature comparable to the midplane temperature. A more complex photochemical prescription may be needed to reproduce the data optimally.

The vertical structure of HCN in the \citet{ionHCN} and Cleeves (private communication) models is even more complicated than in \citet{Wal10}. While highly dependent on the UV field, HCN is predicted to exhibit a double-layered vertical structure caused by chemical and photo-reactions which are dependent on the temperature and ionization levels through the disk. Typical initial fractional abundances for HCN are around $10^{-8}$, which is significantly lower than our measured abundance. Part of the discrepancy between this predicted value and our best fit fractional abundance of $10^{-6.7_{-0.4}^{+0.6}}$ is likely due to non-LTE effects. The LIME model, however, indicated only a factor of two difference in flux between the LTE and non-LTE versions of the best-fit temperature and density structure of the disk, providing tentative evidence that the abundance of HCN may in fact be higher than predicted in this system. A model incorporating several spectral lines, and fully accounting for any NLTE effects, would be useful to break the degeneracy between the temperature and density structure to determine whether the fractional abundance of HCN does in fact differ significantly from the predicted values. 

\subsection{Characterization of Ionizing Radiation Level}\label{sec:rad}

\citet{ionHCN} modeled the emission from a protoplanetary disk irradiated by a nearby O star and compared the gas line emission to that from an isolated disk. They found that while most line emission from the irradiated disk showed higher peak values due to the warmer disk, HCO$^+$, which traces the cold, dense areas of the disk, did not. Thus, the ratio of the HCN/HCO$^+$ peak intensities can be used to roughly characterize the level of external irradiation, with HCN/HCO$^+>1$ indicating an irradiated disk and HCN/HCO$^+<1$ characteristic of an isolated disk. They also predicted that transitions of CO, $^{13}$CO, C$^{18}$O, CI, HCO$^+$, HCN, and CN would be observable at the distance of Orion, with CS and C$_2$H possibly observable in larger, more massive disks. 

We measure the ratio of HCN/HCO$^+$ peak flux in the d216-0939 disk to be $0.58\pm0.04$ (see Section~\ref{res}), comparable to the isolated model from \citet{ionHCN}. This result is not surprising given the location of the disk (relatively isolated and 1.6\,pc projected distance north of $\theta^1$ Ori C, in the outskirts of the Orion Molecular Cloud 2). Our detection of CS \citep[and that of][also part of this large survey of Orion]{ALMABinary} supports their prediction that CS would be observable in larger, more massive disks, since d216-0939 is the largest and one of the most massive disks observed in Orion \citep{MassiveOrion}. We also note that our HCN (4--3) maps are free of cloud contamination, indicating that HCN may be a good disk tracer for observations in regions with contamination in CO or HCO$^+$, especially for disks around intermediate-mass stars.

\subsection{Dynamical Mass of a Pre-Main Sequence Star}\label{sec:mass}

While mass is a fundamental stellar parameter that determines a star's evolutionary path through the HR diagram, theoretical models of pre-main sequence evolution are limited by a lack of understanding of accretion, magnetic fields, and rotation. An accurate determination of the mass of a pre-main sequence star can help to calibrate pre-main sequence evolutionary tracks and thereby determine the ages of host clusters such as the ONC. Spatially resolved molecular line observations of circumstellar disks at millimeter wavelengths provide a powerful diagnostic of the stellar mass \citep[e.g.,][]{IDL2,cze15,cze16}. As discussed in Section \ref{sec:gasmod}, we assume a Keplerian velocity profile in our fit to the molecular line data. The relevant unknowns in calculating this profile are the distance to the system, which affects the conversion of angular to spatial scales, and the inclination of the disk to our line of sight, which affects the projection of the rotational velocity onto the line of sight. If the disk is spatially resolved, the inclination is constrained by the ratio of major to minor axes of the elliptical image (assuming a circular disk). The 1.7\% uncertainty in the distance to Orion \citep[$414\pm7$,][]{OD4} translates directly to an additional 1.7\% uncertainty on the estimated mass.

\begin{figure}
\begin{center}
\includegraphics[width=\columnwidth]{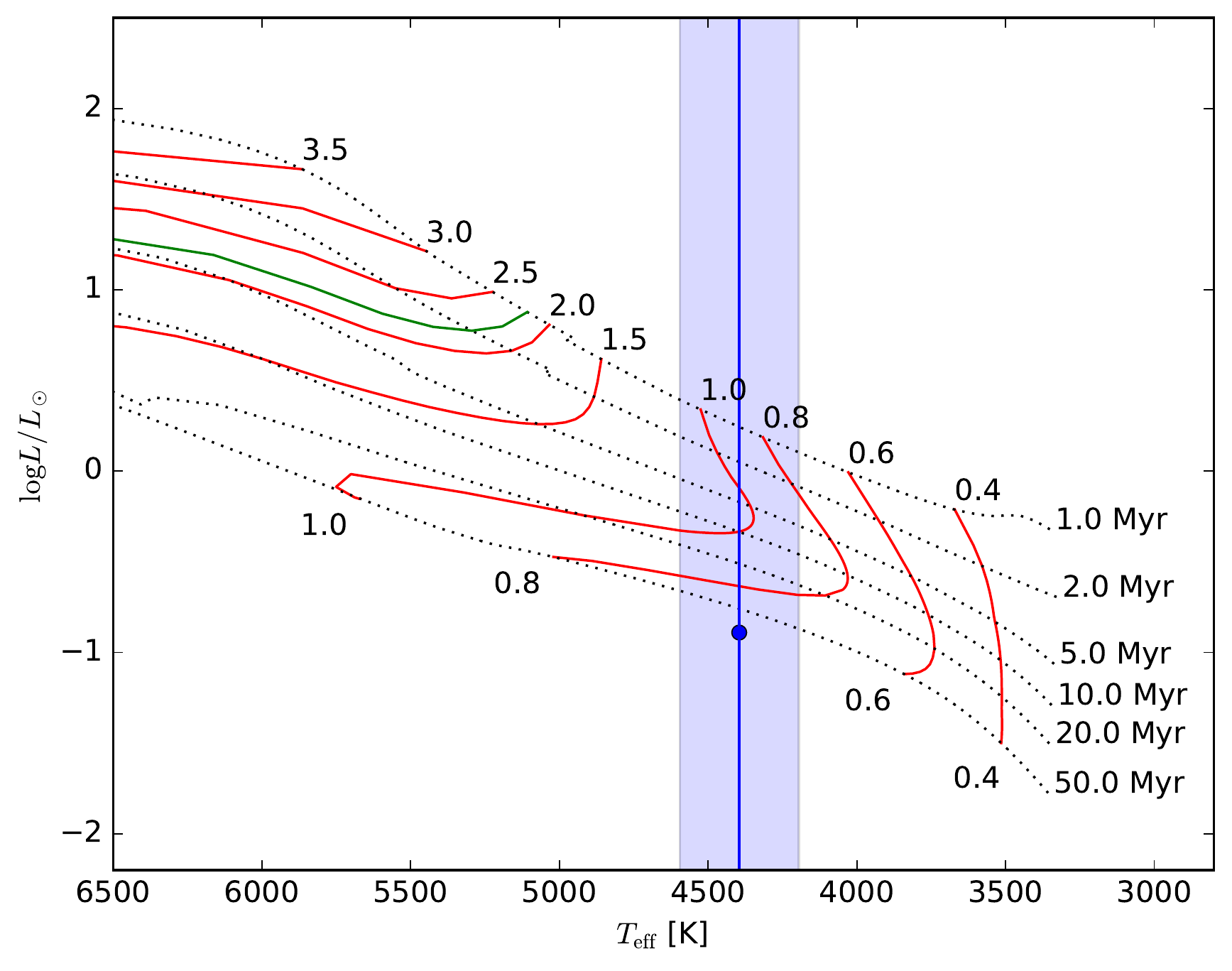}
\caption{Pre-main-sequence tracks from \citet{GFisochrones}. Those shown in red are for context and are labeled with the corresponding stellar mass, in units of $M_\odot$. The green track corresponds to a stellar mass of $2.17\,M_\odot$, the weighted mean of our best fit masses. Isochrones are shown by black dotted lines. The vertical blue line and shaded area indicates the effective temperature and uncertainty of d216-0939, as derived from the spectral type. The blue dot shows the temperature and luminosity (using a bolometric correction to reddening-corrected photometry) reported by \citet{SpT} (ID number 676).}
\label{fig:isochrones}
\end{center}
\end{figure}

Using our dynamical mass measurement and an effective temperature, we can determine an age and luminosity of the star-disk system via a set of theoretical pre-main sequence evolutionary tracks. We first use the spectral type of the star, K5 \citep[determined spectroscopically by][ID number 676, with an uncertainty of $\pm0.5$ for types earlier than K9]{SpT}, to estimate an effective temperature of 4400 K \citep{SpT,Teff1,Teff2,Teff3}. We then place the star on a set of theoretical pre-main-sequence tracks using the estimated temperature and our best fit M$_*$ of $2.17\pm0.07$\,$M_\odot$. The location at which the effective temperature intersects with the stellar mass provides an estimate of the age of the system. Given the stated uncertainty of the spectral classification, we estimate an error in the temperature of $200~\mathrm{K}$. Figure \ref{fig:isochrones} shows evolutionary tracks presented by \citet{GFisochrones} along with our assumed effective temperature. The 2.17\,$M_\odot$ evolutionary track does not intersect with the assumed effective temperature within $1\,\sigma$. This discrepancy could be due to uncertainties in the determination of the spectral type of the star or inaccuracies in the evolutionary tracks. Observations by \citet{HST} using HST showed the disk to be inclined at approximately the flaring angle of the outer disk, causing a large reflection nebula to the east and extinction of the star caused by the disk. The radiative transfer through the disk atmosphere is likely to affect both the observed luminosity and the spectral classification. 

This hypothesis is supported by the extremely low luminosity (blue point in Figure \ref{fig:isochrones}) and thus old age (66 Myr) of the system derived by \citet{SpT}, using a bolometric correction to the reddening-corrected photometry. This age is about 1.5 orders of magnitude older than their own derived age of the Orion Nebula Cluster ($<$1 Myr with a 2 Myr spread). Our dynamical mass measurement, combined with the spectral type and PMS evolutionary tracks, solidly limit the age of the system to $\lesssim1-2$\,Myr. Furthermore the associated HH 667 E and HH 667 W objects and the possible outflow feature shown in Figure \ref{fig:out} suggest that the star may even be younger than 1\,Myr. In order to raise the luminosity of the star up to the 1\,Myr isochrone using the observations of \citet{SpT}, $\sim3-4$ magnitudes of extinction (in the observed $I_C$ band) or reddening (in the calculated $A_V$ correction) must be accounted for. More accurate spectral classification of the star and luminosity measurements, taking into account the dust radiative transfer, are required to better characterize the age of the system.

Another possible explanation for the mass-temperature discrepancy is that the star could be a tight $\sim$equal-mass binary of two $1.1\,M_\odot$ stars which, at 1--2\,Myr, would be consistent with the observed spectral type. We can use the luminosity of a single $2.2\,M_\odot$ star and two $1.1\,M_\odot$ stars to calculate the black-body equilibrium temperature at 150\,au of 38\,K and 34\,K, respectively. Both of these values are between our best fit values for $T_{mid,150}$ and $T_{atm,150}$. We are unable to distinguish between these two equilibrium temperatures as the difference between the two, $\sim10\%$, is about the size of our uncertainty in the atmospheric temperature. A tight binary system would likely clear an inner gap due to tidal interactions with the disk and, when detected, tend to be bright in the millimeter \citep{Har12}. Our observations do not show an inner gap. The slight flux deficit in the HCN (4--3) emission from the inner disk (seen in Figure \ref{fig:moms}) is replicated by our model and is likely chemical in nature. Higher resolution observations could possibly resolve a gap if one exists. We also noted a blue-shifted possible outflow lobe and an associated bow feature $\sim6''$ to the south-east of the disk as seen in Figure \ref{fig:out}. If confirmed to be associated, the misalignment of the lobe with the rotation axis of the disk (by $\sim50^\circ$) could indicate complicated dynamics in the disk, possibly caused by a close in companion. A detailed investigation of this feature is beyond the scope of this work. Again, further observations are required to confirm or refute this binary hypothesis.

\subsection{High Velocity Asymmetry}\label{sec:hvf}

Perhaps the most surprising feature in the data set is the excess of high velocity emission in CO (3--2) and HCO$^+$ (4--3). This feature is clearly visible in the HCO$^+$ (4--3) emission as an additional peak in flux located near the star and along the disk major axis, blue-shifted by approximately 6\,km\,s$^{-1}$ relative to the systemic velocity (see Figures \ref{fig:PV} and \ref{fig:HCOpres}). The feature exceeds $5\,\sigma$ significance in three channels in the HCO$^+$ (4--3) cube. Residuals from fitting a symmetric model to the CO (3--2) emission showed the same feature although more broadly dispersed in the spectral domain, likely due to the larger optical depth of the CO (3--2) line. We attribute this feature to emission originating from the interior regions of the disk and not to cloud contamination for two reasons. First, the feature is localized precisely to the position of the star, where cloud contamination would be distributed across the image. Second, the feature is significantly offset in velocity from the cloud contamination in channels near the systemic velocity. 

To measure the velocity of the feature relative to the star, we fit a Gaussian to the spectrum generated (using the MIRIAD task \texttt{imspec}) from the residual emission. The peak is located at $-6.0\,\mathrm{km\,s}^{-1}$ in HCO$^+$ and $-6.1\,\mathrm{km\,s}^{-1}$ in CO (relative to the systemic velocity), which are consistent to within the channel spacing of 0.4\,km\,s$^{-1}$. Fitting an elliptical Gaussian to the residual visibilities yields a positional offset of $0\farcs11\pm0\farcs04$ north of star at a PA of $-37^\circ\pm9^\circ$, corresponding to a projected linear offset of $50\pm20$\,au. Using the best fit PA and inclination of the disk from Table~\ref{tab:bfp}, this corresponds to a deprojected orbital radius of $57\pm22$\,au. We can then compare the observed velocity of the peak to a Keplerian orbit at the measured position offset. Using the best-fit parameters for the stellar mass and inclination the predicted line of sight velocity of an object in Keplerian orbit at the measured positional offset is $-5.3\pm1.0$\,km\,s$^{-1}$, which agrees well with the measured velocity of the feature. We therefore conclude that the emission is caused by a local density and/or temperature enhancement in the inner disk rather than a separate dynamical process that would deviate from a Keplerian velocity profile such as a jet or outflow \citep[e.g.][]{Podio2015,Tafalla2010}. Such features would be expected to occur at a position angle corresponding to the minor axis of the disk (83$^\circ$), which is inconsistent with the observed position angle of the feature, and would not be expected to exhibit velocities consistent with Keplerian rotation. The agreement of the velocity of the feature and the Keplerian velocity at the deprojected position suggest that the slight difference between the position angle of the feature and the disk major axis is merely a projection effect due to the inclined viewing geometry.

Since the feature is spatially unresolved, we cannot determine its structure. We can, however, use the spectral extent to estimate the radial extent or temperature of the feature, and the integrated flux to estimate the mass of the emitting gas. Based on the Gaussian fit to the spectrum of the feature, we derived a FWHM of 0.94\,km\,s$^{-1}$ and 5.0\,km\,s$^{-1}$ for the HCO$^+$ (4--3) and CO (3--2) emission, respectively. While the broadening in CO (3--2) is likely caused by optical depth, the broadening in the optically thin HCO$^+$ line is more likely physical in nature. This could be caused by a 20\,au radial spread (assuming circular Keplerian orbits) or a 60$^\circ$ azimuthal spread. Assuming only thermal broadening results in a gas temperature of 560\,K. Assuming that the emission is optically thin, we also determined the density enhancement needed to produce such a feature by comparing the peak flux (from the Gaussian fit) to the flux of our best-fit model at the same velocity. We derived a $10\times$ enhancement in density from HCO$^+$ (4--3). The corresponding flux ratio for CO (3--2) is $2\times$, although since the gas is optically thick this value is a lower limit to the density enhancement. 

Finally, we estimated a gas mass of the feature assuming optically thin HCO$^+$ (4--3) emission. We determined the integrated flux in the line, $28\pm10$\,mJy km s$^{-1}$, using the MIRIAD task \texttt{cgcurs} to measure the intensity within the $3\,\sigma$ contour in the zeroth moment map of the residuals in channels corresponding to velocities $\leq-4.3$ km\,s$^{-1}$ relative to the systemic velocity. Using Equation \ref{eq:f2m}, the best-fit HCO$^+$ abundance and atmospheric temperature of the disk (see Equation \ref{eq:Tstruct} and Table \ref{tab:bfp}) at the radius of the feature, we calculate a gas mass of $1.9\pm1.0\,M_\mathrm{Jupiter}$. Using the midplane temperature results in a mass of $1.9\pm1.0\,M_\mathrm{Jupiter}$. Using the temperature calculated assuming thermal broadening alone results in a mass of $8\pm4\,M_\mathrm{Jupiter}$. These masses are highly dependent on the uncertain abundance of HCO$^+$ relative to H$_2$, although our uncertainty on the fitted abundance (which is itself dependent on the fixed disk mass) is considered in the above calculations. If the emission is instead optically thick, as might be the case for a protoplanet envelope, these mass estimates should be considered lower limits.

Deviations from axisymmetry have been observed in several circumstellar disks with ALMA, although primarily in dust rather than gas \citep[e.g.,][]{cas13,mar13,per14}. \citet{Dutrey2014} and \citet{Tang2016} observed a CO clump in the disk of GG Tau that presents some similar features, although that system is more dynamically complex, featuring a hierarchical triple with a large gap in the disk and streamer-like features across the gap, while d216-0939 shows no evidence for an inner gap and is otherwise well described by Keplerian rotation in an unperturbed gas disk. There are several possible explanations for deviations from axisymmetry in a gas disk. One is a recent collision between Mars-size bodies, similar to the proposed explanation for the mid-IR and CO asymmetry in the debris disk around $\beta$ Pictoris described by \citet{betaPicNat} and \citet{betaPicSci}, respectively. Multi-wavelength observations of the continuum which resolve this feature could tell us about the grain sizes present in the feature and could help determine the dynamical process causing the enhancement. \citet{betaPicNat} and \citet{betaPicSci} also suggest that the asymmetry could be caused by collisions between particles trapped in mean motion resonance by a giant planet.

Another possibility is a stellar flyby which could disrupt the disk \citep{flyby}. It is not likely that such an encounter is the cause of this particular feature as it is located in the inner disk and a stellar flyby should primarily perturb the outer disk. The nearest star is $\sim4''$ (1600 au) to the south west and could be a distant companion, though the association has not been confirmed. 

A third possible cause is zonal flows caused by interactions between disk material and a magnetic field. This magneto-rotational instability (MRI) causes pressure and density fluctuations which can cause asymmetries in emission \citep{asym}. These density enhancements are thought to be regions which could promote grain growth and planetesimal formation \citep{zonalFlows}. Our observations have insufficient angular resolution to compare the structure of the observed feature with theoretical predictions, but the appearance of the feature is at least superficially consistent with models of gas overdensities due to zonal flows. The temperature derived from the spectral width of the feature assuming only thermal broadening (560\,K) is inconsistent with temperature enhancements caused by hydrodynamic vortices or zonal flows (of order $\sim$few\% which equates to a few 10s of K, depending on the details of the equation of state and local perturbations of the disk vertical structure; R. Nelson, private communication). The derived azimuthal broadening assuming that the linewidth is due to azimuthal extent ($60^\circ$) is however consistent with long lived vortices \citep{Val-Borro2007,Regaly2012}. \citet{HST} reported large scale asymmetry in the scattered light images from HST, with the northern portion of the disk $\sim50\%$ larger than the southern portion. In contrast, observations of the 850\,$\mu$m continuum emission by \citet{ALMAOri} show that the southern portion of the disk is slightly brighter and our analysis of the gas line emission also shows this. It may be possible that the small-scale asymmetric feature could be causing these large-scale asymmetries by exciting spiral arms or trapping dust \citep[e.g.][]{Zhu15,Zhu14} though higher resolution observations are needed to investigate these effects.

Another possible explanation for the high-velocity asymmetry is planet formation in the disk. \citet{GGP} simulated observations of planet formation in a gravitationally unstable disk, using HCO$^+$ as a tracer of dense gas due to its high critical density. They conclude that HCO$^+$ (7--6) can be used to trace gas giant planet formation through gravitational instabilities and the signature would be observable with ALMA. The high-velocity feature we see in our observations is similar to the planet formation signature they describe, though in a lower $J$ transition of HCO$^+$. The d216-0939 disk, however, is not nearly massive enough to exhibit gravitational instability ($\gtrsim 10\%$ of host star mass). The Toomre $Q$ parameter \citep{toomreQ} calculated using the temperature profiles from our best-fit models never drops below 40, indicating that the disk is likely gravitationally stable throughout. While it is unlikely that the feature is due to gravitational instability, chemical signatures of planet formation in gravitationally stable disks have also been theoretically predicted by \citet{cle15b}. They predict that local heating by luminous young planets undergoing accretion can affect the chemistry of the disk in the region near a forming protoplanet, causing it to shine particularly brightly in HCN and its isotopologues. We do not observe the asymmetry in the HCN (4--3) line, although the non-detection may be due to limitations in the sensitivity of our observations. While \citet{cle15b} do not specifically investigate the signatures of a planet on the HCO$^+$ emission from the disk, it is at least plausible that the predicted local heating could also show up as an enhanced molecular signature in HCO$^+$. The temperature derived from the spectral width of the feature (560\,K) is inconsistent with gas in the surrounding disk, heated by the nearby protoplanet \citep[$\sim30-60$\,K][]{cle15b}. Thus, if the feature is associated with a forming planet, the line width is likely kinematic rather than thermal in nature. Higher-sensitivity observations of HCN and other molecular lines in the d216-0939 disk could indicate whether the chemistry of the region has been altered by an embedded protoplanet. Follow-up observations using ALMA's long baselines would also make it possible to study the radial and azimuthal structure of the feature and compare it with theoretical models of zonal flows, stellar flybys, and the envelopes of forming planets to disentangle the potential explanations for this unusual feature.

\section{Conclusions}
\label{summary}

We have presented spatially and spectrally resolved observations of the HCO$^+$ (4--3), CO (3--2), HCN (4--3), and CS (7--6) molecular line emission from the protoplanetary disk d216-0939 located in the Orion Nebula Cluster (ONC). We used an affine invariant MCMC algorithm to fit a simple model of gas in Keplerian rotation about the central star to the molecular line data, allowing us to characterize the mass of the central star and the temperature and density structure of the gas disk. 

The best-fit parameters are consistent with properties derived for typical massive disks in Taurus and Ophiuchus. These results demonstrate that disks in the ONC, a high-mass star forming region, which are far enough away from $\theta^1$ Ori C not to be externally photoevaporated, can exhibit significant planet-forming potential. Our best-fit HCO$^+$ abundance is consistent with chemical simulations and ISM measurements, while the derived HCN abundance is roughly an order of magnitude larger than simulations and ISM measurements. While non-LTE effects can explain some of the difference in the abundance structure, they do not readily account for the full difference. Future studies at higher angular resolution incorporating coordinated fits to a multi-line data set are needed to investigate the difference conclusively. 

We measure the dynamical mass of the central star to be $2.17\pm0.07\,M_\odot$, which is inconsistent with the spectral type of K5 measured by \citet{SpT}. This discrepancy is possibly due to radiative transfer through the surface layers of the highly inclined disk, or a tight $\sim$equal mass binary. It is possible that future work with more detailed spectroscopic modeling could provide an improved estimate of the stellar temperature and look for signs of binarity, allowing our mass measurement to be used to calibrate the pre-main sequence evolutionary tracks, or to improve the age estimate for the system. 

We serendipitously uncovered a surprising and intriguing feature in the molecular line data as well. Excess high-velocity blue-shifted emission was detected, though its structure could not be spatially resolved, in HCO$^+$ (4--3) and CO (3--2). The spatial and spectral position of this feature are consistent with a local temperature and/or density enhancement in the gas in Keplerian orbit about the central star at a distance of $60\pm20$\,au, far interior to the $\sim$500\,au outer extent of the HCO$^+$ and HCN emission. Using the integrated flux of the feature in HCO$^+$ (4--3), we estimate that the mass of the gas clump is at least $1.8-8\,M_\mathrm{Jupiter}$, depending on the assumed temperature. The characteristics of the feature are reminiscent of theoretical predictions of chemical signatures of forming protoplanets embedded in their gas disks \citep{GGP,cle15b}, although with the existing data we cannot yet distinguish between this explanation and something else, e.g. a localized hydrodynamic vortex. Follow-up observations at higher angular resolution and in multiple molecular tracers will determine whether the chemical and spatial signature matches that of a protoplanet forming within the gas disk. 

\acknowledgments
We thank the referee for their thoughtful comments and L. Ilsedore Cleeves for her helpful advice on disk chemistry. The authors thank Wesleyan University for computer time supported by the NSF under grant number CNS-0619508 and CNS-0959856. This work made use of molecular data made available by \citet{moldat} and Astropy, a community-developed core Python package for Astronomy \citep{astropy}. S.F. acknowledges support from the Connecticut Space Grant Consortium. J.P.W. acknowledges funding from NASA grant NNX15AC92G. This work makes use of the following ALMA data: ADS/JAO.ALMA\#2011.0.00028.S. ALMA is a partnership of ESO (representing its member states), NSF (USA) and NINS (Japan), together with NRC (Canada) and NSC and ASIAA (Taiwan), in cooperation with the Republic of Chile. The Joint ALMA Observatory is operated by ESO, AUI/NRAO and NAOJ. The National Radio Astronomy Observatory is a facility of the National Science Foundation operated under cooperative agreement by Associated Universities, Inc. 

{\it Facilities:} \facility{ALMA}

{\it Software:} CASA \citep{CASA}, MIRIAD \citep{miriad}, LIME \citep{LIME}, emcee \citep{emcee}, corner.py \citep{corner}, Astropy \citep{astropy}

\bibliographystyle{apj}
\bibliography{bib}

\end{document}